\documentclass[10pt, oneside, twocolumn]{article}

\usepackage{geometry} 
\usepackage{graphicx}
\usepackage{amssymb}
\usepackage{amsmath}
\usepackage{empheq}
\usepackage{physics}
\usepackage{tensor}
\usepackage{csquotes}
\usepackage{sectsty}
\sectionfont{\small}
\usepackage[colorlinks=true, allcolors=blue]{hyperref}
\usepackage{xcolor}
\usepackage[normalem]{ulem}
\usepackage{tikz}
\usetikzlibrary{decorations.pathreplacing,calc,tikzmark}
\usepackage[font=small,labelfont=bf]{caption}
\usepackage{cite}
\usepackage{authblk}

\usepackage{titlesec}
\titleformat{\subsection}[runin]{\bfseries}{}{0em}{}[]
\makeatletter
\renewcommand{\maketitle}{\bgroup\setlength{\parindent}{0pt}\fontsize{15}{19.2}\selectfont

\begin{flushleft}
  \textbf{\@title}
  
  \@author
\end{flushleft}\egroup
}
\makeatother

\title{Simultaneous Brillouin and piezoelectric coupling to a high-frequency bulk acoustic resonator}

\author[1,2,$\dagger$,*]{Taekwan Yoon}
\author[1,2,*]{David Mason}
\author[1,2]{Vijay Jain}
\author[3]{Yiwen Chu}
\author[1,2]{Prashanta Kharel}
\author[4]{William H. Renninger}
\author[5]{Liam Collins}
\author[1,2]{Luigi Frunzio}
\author[1,2,**]{Robert J. Schoelkopf}
\author[1,2,***]{Peter T. Rakich}

\affil[1]{Departments of Applied Physics and Physics, Yale University, New Haven, CT 06520, USA}
\affil[2]{Yale Quantum Institute, New Haven, CT 06520, USA}
\affil[3]{Department of Physics, ETH Zurich, 8093 Zurich, Switzerland}
\affil[4]{Institute of Optics, University of Rochester, Rochester, NY 14627, USA}
\affil[5]{Center for Nanophase Materials Sciences, Oak Ridge National Laboratory, Oak Ridge, TN 37831, USA}

\affil[$\dagger$]{Corresponding author: taekwan.yoon@yale.edu}
\affil[*]{These two authors contributed equally}
\affil[**]{robert.schoelkopf.yale.edu}
\affil[***]{peter.rakich@yale.edu}

\begin{document}

\twocolumn[

\begin{@twocolumnfalse}
\maketitle

\begin{abstract}
Bulk acoustic resonators support robust, long-lived mechanical modes, capable of coupling to various quantum systems.
In separate works, such devices have achieved strong coupling to both superconducting qubits, via piezoelectricity, and optical cavities, via Brillouin interactions.
In this work, we present a novel hybrid microwave/optical platform capable of coupling to bulk acoustic waves through cavity-enhanced piezoelectric and photoelastic interactions.
The modular, tunable system achieves fully resonant and well-mode-matched interactions between a 3D microwave cavity, a high-frequency bulk acoustic resonator, and a Fabry Perot cavity.
We realize this piezo-Brillouin interaction in x-cut quartz, demonstrating the potential for strong optomechanical interactions and high cooperativity using optical cavity enhancement.
We further show how this device functions as a bidirectional electro-opto-mechanical transducer, with transduction efficiency exceeding $10^{-8}$, and a feasible path towards unity conversion efficiency.
The high optical sensitivity and ability to apply large resonant microwave field in this system also offers a new tool for probing anomalous electromechanical couplings, which we demonstrate by investigating (nominally-centrosymmetric) CaF\textsubscript{2} and revealing a parasitic piezoelectricity of 83 am/V.  Such studies are an important topic for emerging quantum technologies, and highlight the versatility of this new hybrid platform.
\end{abstract}

\end{@twocolumnfalse}]

\section{Introduction}

Acoustic resonators with long-lived modes have played an important technological role, in systems ranging from ultrastable oscillators to micro-electro-mechanical devices.
For quantum applications, acoustic devices offer the promise of flexibly linking disparate quantum systems, through a wide variety of coupling pathways\cite{kurizki_quantum_2015}.
Mechanical resonators have achieved quantum-coherent coupling to systems ranging from optical and microwave cavities, to solid state defect centers, atomic ensembles, and superconducting qubits\cite{chan_laser_2011,oconnell_quantum_2010,verhagen_quantum-coherent_2012,teufel_sideband_2011, maity_coherent_2020, whiteley_spinphonon_2019, chu_quantum_2017}.
These advances relied on high quality-factor (Q) mechanical resonators, building on phononic engineering and low material dissipation at cryogenic temperatures\cite{beccari_strained_2022, maccabe_nano-acoustic_2020}.  
Among mechanical systems, high-overtone bulk acoustic resonators (HBARs), where high-frequency elastic standing waves are formed between polished surfaces of a crystalline substrate, offer a promising platform for achieving high acoustic Q-factors\cite{galliou_extremely_2013}.
By design, these modes live primarily in the bulk, avoiding the loss typically associated with surface imperfections. Moreover, acoustic loss from phonon-phonon scattering plummets at cryogenic temperatures\cite{galliou_extremely_2013, liekens_attenuation_1971, thaxter_phonon_1966}, offering an ideal environment for long-lived acoustic modes.

To achieve strong optical interactions with these acoustic modes, one can exploit phase-matched photoelastic coupling, known as Brillouin scattering\cite{chiao_stimulated_1964, renninger_bulk_2018, bourhill_generation_2020, kharel_high-frequency_2019}.
This three-wave interaction can yield strong optomechanical coupling 
to GHz-frequency acoustic modes, particularly when the process is enhanced by an optical cavity.
One can also couple microwave fields directly to the acoustic modes via resonant piezoelectric interaction.
Using these interactions, in separate systems, bulk acoustic resonators have achieved strong coupling to both superconducting circuits and optical cavities\cite{chu_quantum_2017, chu_creation_2018, kharel_high-frequency_2019, kharel_multimode_2019}.

Here we present, for the first time, a system integrating simultaneous piezoelectric and Brillouin coupling to a GHz bulk acoustic resonator, in a hybrid system containing both a microwave and an optical cavities.
To demonstrate the mechanism behind this piezo-Brillouin interaction, we first present a single-pass configuration (i.e. without the benefit of optical cavity enhancement), using a microwave cavity to drive GHz phonon modes in x-cut quartz, and reading out scattered optical photons.
We then incorporate an optical cavity to realize a fully resonant operation where both electromechanical and optomechanical coupling mechanisms are cavity-enhanced, boosting interaction rates by orders of magnitude. The use of a modular Fabry-Pérot cavity offers frequency tunability to meet this multiply-resonant condition, while maintaining low loss optical modes.

We observe strong optomechanical interactions (characterized by a cooperativity exceeding unity), even reaching strong coupling.
Along with competitive coupling efficiencies 
and robust thermal properties,
this reveals the potential of this hybrid platform for microwave-optical transduction with quantum applications\cite{zeuthen_figures_2020, lauk_perspectives_2020, chu_perspective_2020, andrews_bidirectional_2014, vainsencher_bi-directional_2016,balram_coherent_2016,jiang_efficient_2020, forsch_microwave--optics_2020, han_cavity_2020, mirhosseini_superconducting_2020, rueda_efficient_2016, hease_bidirectional_2020,holzgrafe_cavity_2020, shao_microwave--optical_2019, fan_superconducting_2018}.
Realizing efficient quantum state conversion is of critical importance for optically linking superconducting quantum devices, and remains an outstanding goal for the field.
Here we present bidirectional conversion with transduction efficiency comparable to other developing piezo-optomechanical platforms \cite{blesin_quantum_2021, vainsencher_bi-directional_2016,balram_coherent_2016,jiang_efficient_2020,forsch_microwave--optics_2020,han_cavity_2020,mirhosseini_superconducting_2020}, and we outline a feasible path for improvement.

We further demonstrate that the optical performance of this system offers a highly sensitive method for detecting piezoelectrically-driven phonons, offering a new tool for materials analysis.
We illustrate this by interrogating CaF\textsubscript{2},  revealing an anomalous piezoelectric coupling, which may indicate surface damage or lattice defects in the bulk\cite{robinson_electrical-mechanical_1978, sharma_possibility_2007,  aktas_piezoelectricity_2021}.
Probing such effects in technologically-relevant materials (e.g. sapphire or diamond) can offer clues to understanding loss mechanisms for quantum systems\cite{georgescu_surface_2019, maze_properties_2011}.

\section{Demonstration of single-pass piezo-Brillouin spectroscopy}

\begin{figure}[h!]
    \centering   \includegraphics[width=7.5cm]{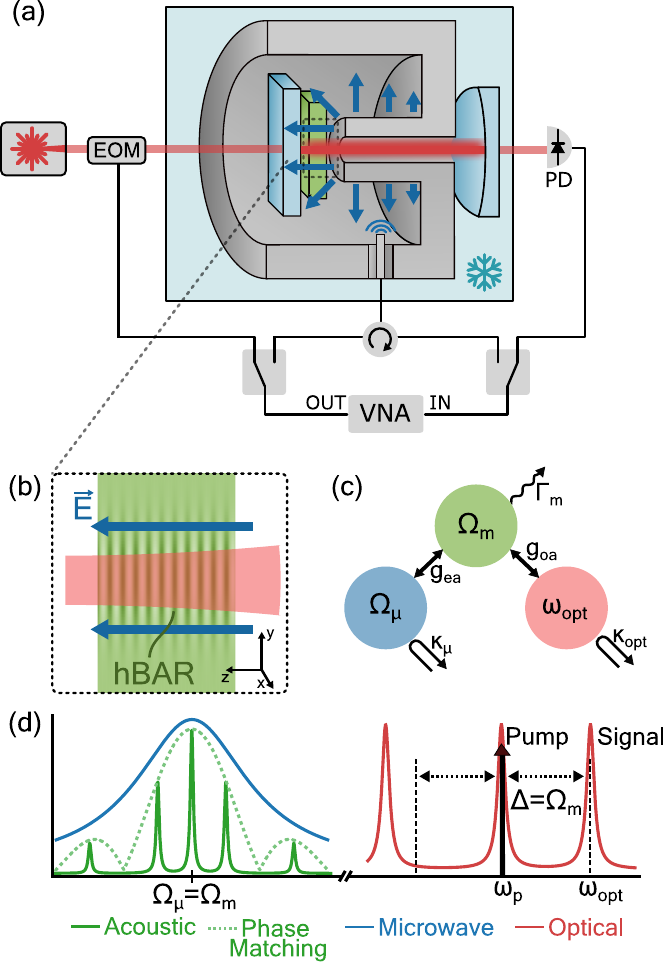}
        \caption{
        \textbf{Hybrid piezo-Brillouin optomechanical system.}
        (a) Cross-section of the hybrid cavity and simplified experimental schematic, showing the acoustic resonator (green), microwave cavity (grey), and optical mirrors (light blue). Dark blue arrows indicate microwave E-field, dark green indicates the longitudinal acoustic mode of the HBAR, and red denotes optical light.
        Microwave and optical signals can be injected by driving the microwave cavity or electro-optic modulator (EOM), respectively.
        Optical signals are detected in transmission (at photodiode, PD), using the pump light as heterodyne local oscillator. Microwave signals can be collected and amplified directly from the cavity. Experiment conducted in He-4 flow cryostat, $T\approx 10$ K. See Supplement for detailed experimental schematic.
        (b) Inset showing overlap between the microwave/optical/acoustic modes
        (c) Simplified illustration of mode couplings and decay pathways for the optical (red), microwave (blue), and acoustic (green) modes.
        (d) Spectral response of microwave (blue), optical (red), and acoustic (green) modes, including the Brillouin phase-matching bandwidth (dashed green). The Stokes and anti-Stokes sideband frequencies are indicated by black dashed lines, symmetrically spaced around the pump ($\omega_\mathrm{p}$, black arrow). Only the anti-Stokes is resonant with an optical mode (at $\omega_\mathrm{opt}$).
        Non-uniform optical mode spacing is caused by the Fresnel reflections at the vacuum/dielectric interface of the crystal within the cavity \cite{kharel_high-frequency_2019}.
        }
        \label{fig1}
\end{figure}

Our hybrid platform (illustrated in Fig.  \ref{fig1}(a)) is designed to achieve fully resonant microwave and optical coupling to an HBAR device.
In contrast to integrated nanomechanical devices, this platform is an assembly of three distinct resonators (acoustic, optical, and microwave).
This modular design enables independent optimization of each resonator, including quality factors and mode-matchings. The optical and microwave frequencies are individually tunable, to allow resonant coupling to virtually any transparent crystalline material.
We demonstrate this hybrid piezo-Brillouin interaction using x-cut quartz -- a piezoelectric material with good acoustic performance.

The platform design centers around the use of an HBAR, that supports longitudinal elastic standing waves confined between the opposing surfaces of a crystalline substrate.  
At modest cryogenic temperatures ($T\sim4$ K), these Fabry-Perot-like acoustic modes can achieve quality factors at GHz (MHz) frequencies in excess of $10^7$ ($10^9$), by working with high purity crystals with smooth surfaces \cite{galliou_extremely_2013, kharel_ultra-high-q_2018}.
While plano-convex crystals offer stable resonances with the highest mechanical Q, here we work with flat, unprocessed substrates for material flexibility and simplified assembly. Acoustic modes of a flat-flat resonator experience diffraction loss, but still reach $Q>10^4$.
The substrate hosts a set of longitudinal modes, spaced by a mechanical free spectral range, $\Delta_\mathrm{m} = v_\mathrm{m}/2L_\mathrm{m}$, determined by the sound velocity, $v_\mathrm{m}$, and substrate thickness, $L_\mathrm{m}$.
Here, we work with a 0.5mm thick x-cut quartz substrate, with acoustic loss rate of $\Gamma/2\pi\approx 500$ kHz and $\Delta_\mathrm{m}/2\pi\approx 5.5$ MHz.

For electromechanical coupling to the HBAR, we rely on resonant piezoelectric interactions with a microwave resonator as shown in Fig. \ref{fig1}(c).
We require a microwave resonator that can match the acoustic frequency (10-50 GHz), ideally with high Q-factor and easy tunability.
A coaxial stub cavity (see Fig. \ref{fig1}a), a common resource for quantum technologies, is well-suited for this task\cite{reagor_quantum_2016}.
In such a resonator, the microwave field is concentrated around a quarter-wavelength central post, allowing for easy adjustment of the resonance frequency via the post length. 
The monolithic cavity geometry eliminates seam losses, which is beneficial in creating high Q cavities. Combined with the use of superconducting materials, the Q-factor of a coaxial stub cavity can exceed $Q_\mu>10^7$\cite{reagor_quantum_2016}.
Here, we work at T = 4K, and choose (non-superconducting) copper for its thermal conductivity and low electrical resistivity, reaching microwave loss rate of $\kappa_\mu/2\pi \approx$ 17 MHz, and corresponding internal Q-factor of $Q_\mu \approx$ 1200. 
Additionally, we make a small (1 mm diameter) aperture through the stub to allow for optical access, which has negligible impact in the Q-factor of the microwave design.

The coaxial microwave cavity is resonantly coupled, via the piezoelectric effect, to an HBAR located above the central post. Electromagnetic simulations (Ansys High Frequency Simulation Software - HFSS) show that the electric field is oriented perpendicular (i.e. the $z$-direction; see Fig. \ref{fig1}b) to the substrate, and is approximately uniform across the substrate thickness.
This field resonantly couples to the longitudinal acoustic mode via the $d_{33}$ element of the piezoelectric tensor. Note that all crystal axis indices are assumed to be in the reference frame of the substrate.
Specifically, the piezoelectric interaction is described by the coupling rate, $\hbar g_{\mathrm{em}} = c_{33} d_{33} \int E_z S_{zz} dV$,
where $S_{zz}=\frac{\partial U_z}{\partial z}$ is the longitudinal strain, expressed in terms of the z-displacement, $U_z$, $E_z$ is the z-component of the microwave electric field, and $c_{33}$ is the relevant stiffness coefficient.
While the acoustic mode oscillates along the $z$-axis with a sub-micron wavelength, the microwave electric field is uniform along the thickness of the crystal.
As a consequence, the electromechanical mode-matching integral vanishes for acoustic modes with even longitudinal indexes.
Odd-indexed modes couple at a rate
\begin{equation}
    g_{\mathrm{em}}=\frac{d_{33}E_0\lambda_\mathrm{m}}{\pi}\sqrt{\frac{\Omega_\mathrm{m} c_{33}A_\mathrm{m}}{2\hbar L_\mathrm{m}}},
\end{equation}
where $E_0$ is the zero-point field strength, $\lambda_\mathrm{m}$ ($\Omega_\mathrm{m}$) is the acoustic wavelength (frequency), $A_\mathrm{m}$ is the effective area of the acoustic mode, and $L_\mathrm{m}$ is the acoustic substrate thickness (see Supplement for derivation).
The interaction strength of the microwave and acoustic resonators can be characterized by the electromechanical cooperativity, $C_\mathrm{em} = \frac{4g_\mathrm{em}^2}{\kappa_\mu \Gamma}$.
From this piezoelectricity and microwave simulations, we estimate the electromechanical coupling in this quartz system to be $g_{\mathrm{em}}/2\pi=298$ Hz, corresponding to $C_\mathrm{em} = 4.18\times 10^{-8}$.
It is important to use cryogenic values for quartz piezoelectricity\cite{tarumi_complete_2007}, as it can vary $\sim$50\% at cryogenic (5K) temperatures.

Optomechanical coupling is realized through Brillouin scattering, which is a three-wave interaction between optical pump/signal fields and an acoustic wave, mediated by photoelasticity\cite{chiao_stimulated_1964, renninger_bulk_2018}.
Brillouin scattering offers a distributed, phase-matched interaction that takes place throughout the bulk of the crystal.
This is advantageous compared with conventional radiation pressure optomechanical coupling, which is relatively weak due to the large acoustic mode volume of HBARs.
The Brillouin interaction is described by the single-photon coupling rate, $\hbar g_{\mathrm{om,0}} =\frac{1}{2}\epsilon_0\epsilon_r^2p_{13}\int S_{zz} E^2_{\mathrm{opt}}dV$, where
$E_\mathrm{opt}$ is the transverse electric field of the optical cavity (assumed to be along $\hat{x}$ here), $p_{13}$ is the relevant photoeleastic coefficient, and $\epsilon_0$ ($\epsilon_r$) is the vacuum (relative material) permittivity\cite{renninger_bulk_2018, van_laer_unifying_2016}.
The mode overlap is maximized when the modes satisfy a phase-matching condition, $2k_\mathrm{opt} \cong q$, where $k_\mathrm{opt}$ ($q$) is the optical (acoustic) wave vector. This phase-matching condition implies that the acoustic wavelength should approximately match half of the optical wavelength in the medium.
This results in non-zero couplings to a small number of acoustic modes centered around the Brillouin frequency ($\Omega_\mathrm{B} = 2\omega_\mathrm{p} n v_\mathrm{m}/c$), where $
\omega_\mathrm{p}$ is the optical pump frequency, $n$ is the refractive index, and $v_\mathrm{m}$ is the acoustic velocity in the substrate.
For x-cut quartz, the Brillouin frequency is $\Omega_\mathrm{m}/2\pi\approx$ 11.4 GHz for $\lambda_\mathrm{opt}$ $\approx$ 1550 nm.
Hereafter, we designate the longitudinal mode best satisfying this phase-matching condition as the primary mechanical mode of interest, with frequency $\Omega_\mathrm{m}$ (see Fig. \ref{fig1}d).
For this phase-matched mode, the Brillouin optomechanical coupling rate is\cite{renninger_bulk_2018, kharel_high-frequency_2019}
\begin{equation} \label{eq_gom}
    g_{\mathrm{om,0}} = \frac{\omega_\mathrm{p}^2n^3p_{13}}{2c}\sqrt{\frac{\hbar}{\Omega_\mathrm{m}\rho A_\mathrm{m} L_\mathrm{m}}},
\end{equation}
where $\rho$ is the density and $c$ is the speed of light.
Note that here we assumed that the acoustic mode area is defined by the optical mode area $A_\mathrm{opt}/2 = A_\mathrm{m}$ (see supplement).
Other acoustic modes are suppressed by a $\mathrm{sinc}^2\left[ (\Omega-\Omega_\mathrm{B})/4\Delta_\mathrm{m}\right]$ envelope, that originates from the phase-matching condition.
Drawing an analogy to cavity-optomechanics\cite{aspelmeyer_cavity_2014}, $g_\mathrm{om,0}$ can be linearized around a strong optical pump, resulting in an enhanced interaction between the acoustic mode and optical signal field.  This enhanced coupling rate is given by $g_{\mathrm{om}}=\sqrt{N}g_{\mathrm{om,0}}$, where $N$ is the number of pump photons in the crystal.

Brillouin scattering measurements are conventionally implemented in a single-pass configuration, where a strong pump can interact with acoustic phonons, scattering into backwards-propagating Stokes and anti-Stokes fields.  In this configuration, a Brillouin-active phonon mode, with population $n_\mathrm{m}$, will scatter incident pump photons to an anti-Stokes sideband with power $P_{\mathrm{sig}} =  (\hbar \omega_\mathrm{p}) C_\mathrm{om}^{\mathrm{sp}} \Gamma n_\mathrm{m}$.
Here, $C_\mathrm{om}^{\mathrm{sp}} = \frac{g_{\mathrm{om}}^2}{\Gamma\tau^{-1}} $ is the single-pass equivalent of a cavity cooperativity\cite{renninger_bulk_2018}, in terms of the group delay $\tau = L_\mathrm{m} n/c$ (see Supplement).
Fig.  \ref{fig2}(d) presents such a Brillouin signal, from piezoelectrically driven phonons, using the previously described microwave cavity and x-cut quartz HBAR.
We sweep a 3.8 mW microwave drive over a $\sim$40 MHz range around $\Omega_\mathrm{B}$ while measuring the coherent back-scattered optical signal, detected as a heterodyne beat note with reflected pump light.
We see a series of acoustic resonances with $\Gamma/2\pi\approx$ 500 kHz and the expected mode spacing $\Delta_\mathrm{m} \approx$ 5.5 MHz. The resonances away from $\Omega_\mathrm{B}$ are suppressed according to the optical phase-matching condition and microwave cavity susceptibility.
For this single-pass measurement, we theoretically estimate the single-photon coupling rate to be $g_\mathrm{om,0}/2\pi=156$ Hz.  For our pump power (P = 67 mW), this gives a field-enhanced $g_\mathrm{om}/2\pi=180$ kHz ($C_\mathrm{om}^{\mathrm{sp}}=1.1\times 10^{-6}$), which predicts the observed signal size within 5 dBm (see Fig. \ref{fig2}d). The discrepancy can be attributed to the uncertainties in material constants at cryogenic temperatures.
While this measurement demonstrates piezo-Brillouin interaction, it is desirable to have higher optomechanical coupling for broader applications, ideally exceeding a cooperativity of unity.
To achieve this, we incorporate an optical cavity into this piezo-Brillouin system.

\section{Fully resonantly-enhanced piezo-Brillouin spectroscopy}

\begin{figure}[h!]
    \centering
        \includegraphics[width=7.5cm]{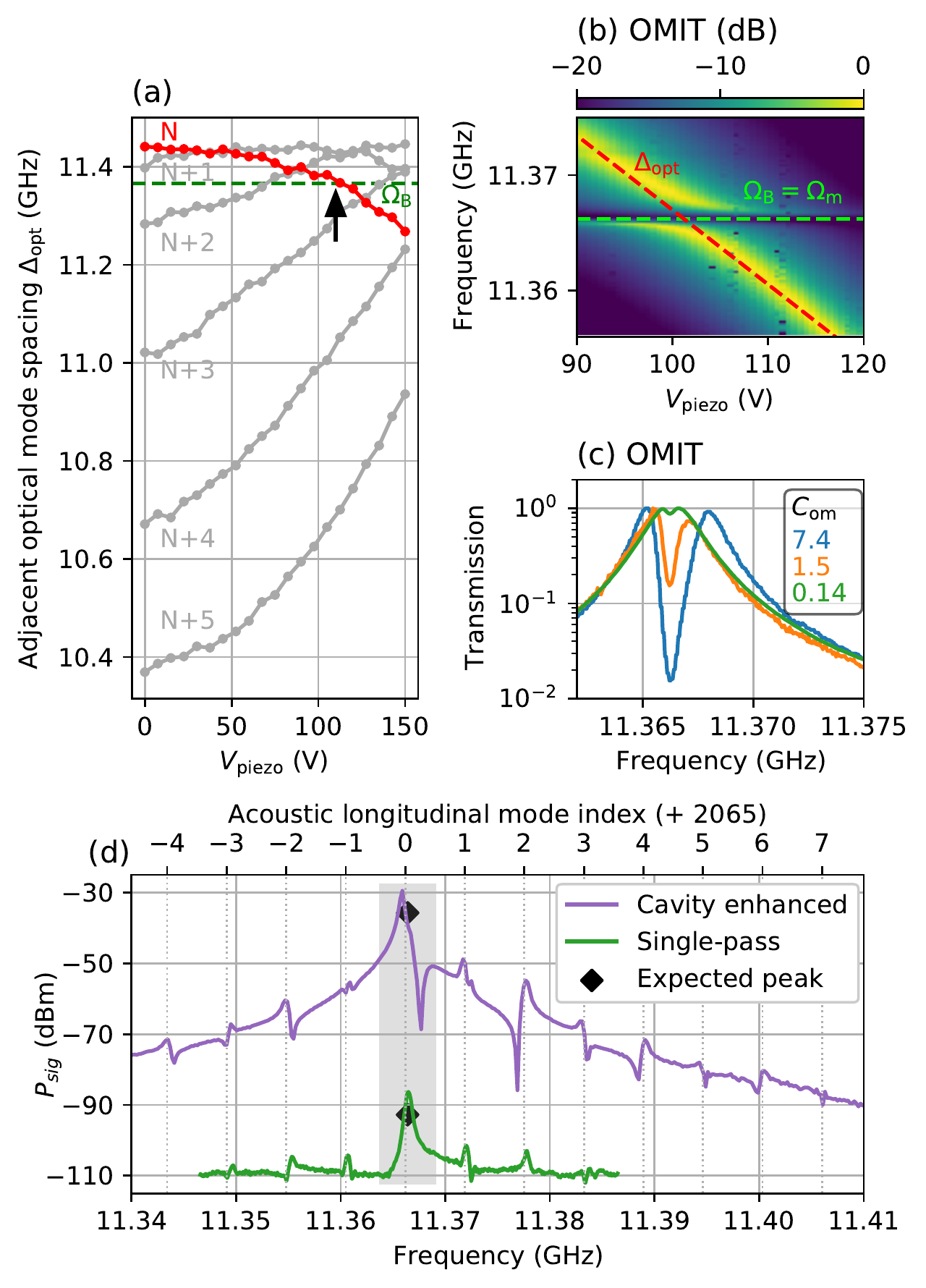}
        \caption{
        \textbf{Enhancing Piezo-Brillouin spectroscopy using resonant optomechanical interactions}
        (a) Optical mode spacing tunability.  For $\lambda_\mathrm{opt}$ between 1545.8-1546.3 nm, we find 6 mode pairs (labelled by lower longitudinal mode index N through N+5) with spacings from 10.4 GHz to 11.5 GHz.  By increasing $V_\mathrm{piezo}$ (thus changing cavity length), individual mode spacings can be varied by up to 500MHz, allowing one mode pair (red) to be matched to the $\Omega_\mathrm{B}$ (green dashed line)
        (b) Normalized OMIT Spectra, as $\Delta_\mathrm{opt}$ is tuned through $\Omega_\mathrm{B}$, using the piezo actuator.  When $\Delta_\mathrm{opt}=\Omega_\mathrm{B}$, an anti-crossing is observed, with a splitting that exceeds the cavity linewidth, indicating the onset of strong coupling.
        (c) Single OMIT Spectra at $\Delta_\mathrm{opt}=\Omega_\mathrm{B}$ with fitted cooperativities ranging from $C_\mathrm{om}=7.4\sim0.14$.
        (d) Single-pass and cavity-enhanced piezo-Brillouin spectroscopy. Sideband power at the swept microwave drive frequency is measured via heterodyne with the pump (1kHz resolution bandwidth). Grey box indicates the main phonon mode of interest, with theoretically predicted peak amplitude shown as black dots. Other acoustic modes are spaced by the expected FSR ($\approx$ 5.5 MHz) and have couplings consistent with mode- and phase-matching.
        To facilitate comparison more easily, the green data has been shifted down in frequency by 9 FSRs to compensate for a $\sim0.5\%$ shift of microwave resonance frequency between cooldowns.
        }
        \label{fig2}
    
\end{figure}

Including an optical cavity into this piezo-Brillouin system allows for resonant enhancements of both the optical pump and signal fields\cite{van_laer_unifying_2016}.
As demonstrated in prior works \cite{kharel_high-frequency_2019, kharel_multimode_2019}, a Fabry-Perot cavity is well-suited to this task, with Gaussian modes that achieve good spacial overlap with highly confined HBAR modes (see Fig. \ref{fig1}a-b).
Compared to the single-pass interaction, the effect of this optical resonator is to increase the optomechanical cooperativity (and thus $P_\mathrm{sig}$) by a factor of $\mathcal{F}^2$, where $\mathcal{F}$ is the cavity finesse, a quantity that indicates the number of round trips before decay (see Supplement).
Specifically, the cavity optomechanical cooperativity is now $C_\mathrm{om}=\frac{4g_\mathrm{om}^2}{\kappa_\mathrm{opt}\Gamma}$ with $g_\mathrm{om}=\sqrt{N_\mathrm{cav}}g_\mathrm{om,0}$, where  $N_\mathrm{cav}$ is now the intracavity photon number and $\kappa_\mathrm{opt}$ is the optical cavity dissipation rate.
Note that, compared to the single-pass coupling rate, $g_\mathrm{om,0}$ acquires a factor of $\frac{L_\mathrm{m}}{L_\mathrm{opt}}$, corresponding to a geometric filling factor between the acoustic and optical cavities.

For our optical cavity geometry, even a modest finesse ($\mathcal{F}>1000$) results in optical linewidths that are orders of magnitude less than the mechanical frequency (i.e. $\kappa/\Omega_\mathrm{m} \sim 10^{-3}$).
In this deeply resolved-sideband regime, it becomes necessary to use separate resonances to enhance the pump and signal, thus imposing an additional requirement that the optical mode spacing closely matches the acoustic frequency ($\Delta_\mathrm{opt}\equiv\omega_\mathrm{opt}-\omega_\mathrm{p}=\Omega_\mathrm{m}$).  

Two features of this hybrid system allow us to meet this requirement.
First, the Fresnel reflections at the vacuum/dielectric interface produced by the HBAR cause the optical mode spacing to be wavelength-dependent. 
For instance, in our hybrid quartz cavity, over a $\sim$1 nm range in pump wavelength, we find 9 optical modes with spacings varying from 10.4 GHz to 11.5 GHz, which allows us to identify several pairs of modes with a frequency spacing that approximately match $\Omega_\mathrm{B}$.
Note that this variable FSR allows selective enhancement of either the Stokes or anti-Stokes process.
While this coarse frequency-matching was possible in prior works\cite{kharel_high-frequency_2019, kharel_multimode_2019}, here we incorporate a piezo actuator (voltage, $V_\mathrm{piezo}$) onto one of the mirrors, to enable more precise control required in this multiply resonant system.
This allows the cavity length to be varied by approximately 500 nm, which can yield $>$ 100 MHz tunability in the optical mode spacing (see Fig. \ref{fig2}a).

This matching of the optical mode spacing to the mechanical frequency enables strong optomechanical interactions, which we illustrate in Fig. \ref{fig2}(b-c).  We lock a laser to the lower frequency (pump) mode, and sweep a sideband over the higher frequency (signal) mode.  When $\Delta_\mathrm{opt}\neq\Omega_\mathrm{m}$, this swept probe simply shows the optical response of the signal mode.  Applying a voltage to the piezo actuator, we can smoothly vary the optical mode spacing from 11.355 GHz to 11.375 GHz (red dashed line in Fig.  2b), revealing an avoided crossing when $\Delta_\mathrm{opt}=\Omega_\mathrm{m}$.  This phenomenon is known as optomechanically induced transparency (OMIT), and is the result of optical interference with a photoelastically-driven motional sideband\cite{weis_optomechanically_2010}. OMIT measurements offer a convenient tool for characterizing the strength of our optomechanical interaction.
Specifically, the depth and width of the OMIT feature offer a direct measurement of the optomechanical cooperativity, with minimal assumptions.  Fig. \ref{fig2}(c) shows OMIT spectra for various pump powers. The extracted cooperatvities scale linearly with intracavity power, as expected, reaching $C_\mathrm{om}$ = 7 ($g_\mathrm{om} = 1.4$ MHz) at a pump power of 110 mW.
Combining this $g_\mathrm{om}$ with the estimated intra-cavity photon number, we can obtain an experimental value for single-photon optomechanical coupling rate, $g_\mathrm{om,0} = 5.28$ Hz, close to the theoretical value, $g_\mathrm{om,0} = 5.89$ Hz.
Fig. \ref{fig2}(b-c) also illustrate that this system approaches the optomechanical strong coupling regime ($2g_\mathrm{om}>\kappa_\mathrm{opt},\Gamma$), visible as the emergence of normal-mode splitting\cite{kharel_multimode_2019}.
Such strong coupling can be of value in hybrid quantum systems, for rapid manipulation of intracavity states.

With this doubly-resonant optical enhancement established, we repeat the measurement from Fig. \ref{fig2}(d), wherein a microwave drive piezoelectrically generates phonons and we measure the optical sideband scattered off a strong pump.
The transmitted optical signal is collected from the cavity and measured via heterodyne beat note with the pump. 
The optical response from the microwave drive ($P_\mu$ = 1 mW) is shown in purple in Fig. \ref{fig2}(d).
The central phase-matched acoustic mode ($\Omega_\mathrm{m}/2\pi$ = 11.366 GHz) is indicated by a gray shaded box.
Comparing the cavity-enhanced and single-pass spectra, we observe a $\sim$60dB increase in scattered optical power, consistent with the predicted $\mathcal{F}^2$ enhancement in cooperativity (for details, see Supplement).

The cavity-enhanced data illlustrates several prominent features that are consistent with the underlying coupling mechanisms. On the tails of the central, phase-matched peak ($\Omega_\mathrm{m}/2\pi$=11.366 GHz), we see the other longitudinal modes, whose motion coherently interferes to yield Fano lineshapes. The apparent coupling rates of these modes are suppressed according to the optical phase-matching condition ($ \mathrm{sinc}^2\left[ (\Omega-\Omega_\mathrm{B})/4\Delta_\mathrm{m}\right]$),
as well as the piezoelectric mode-overlap
(which suppresses coupling to even-index modes).
Away from the central mode, the piezo-Brillouin spectrum is further suppressed by the optical and microwave cavity susceptibilities.
There is also a significant Fano dip associated with the main peak at $\Omega_\mathrm{m}/2\pi$ = 11.366 GHz. This is a result of a residual Pockels-type electro-optic (EO) coupling in quartz\cite{ivanov_direct_2018}, which destructively interferes with the piezo-optomechanical signal.
This Fano shape results in a slightly higher response on the left side of the peak.
The relative contribution of the EO effect to the piezo-Brillouin interaction can be reduced by increasing $g_\mathrm{em}$ or decreasing $\Gamma$.
Alternatively, this effect could be intentionally amplified in a device focused on electro-optic interactions.

\section{Transduction}
\begin{figure}[h!]
    \centering
        \includegraphics[width=7.5cm]{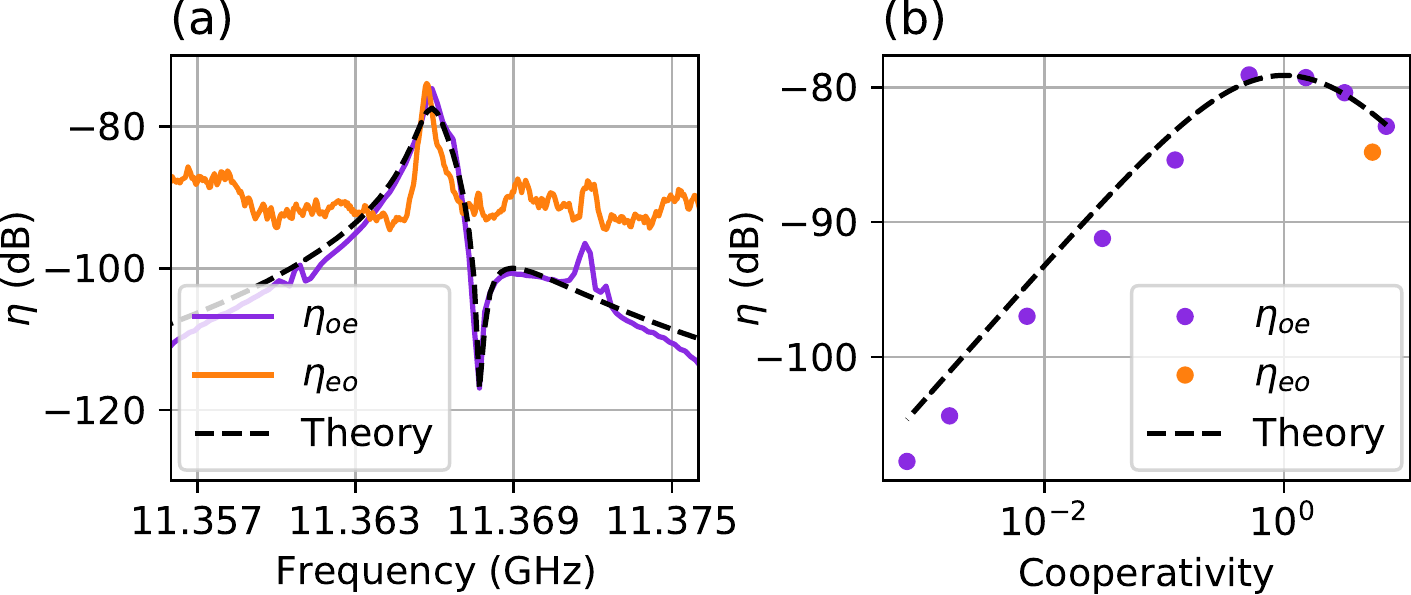}
        \caption{
        \textbf{Bidirectional electro-opto-mechanical transduction}
        (a) Transduction spectra, for both microwave-to-optical ($\eta_{\mathrm{oe}}$, purple, $C_{\mathrm{om}}=1.53$) and optical-to-microwave ($\eta_{\mathrm{eo}}$, orange, $C_{\mathrm{om}}=5.50$). Dashed black line is theoretical prediction of the transduction spectra.
        The elevated noise floor in the $\eta_\mathrm{eo}$ is set by the room temperature amplifiers/detectors.
        (b) Transduction efficiency at phonon frequency ($\Omega_\mathrm{m}/2\pi = 11.3663$GHz) in relation to the optomechanical cooperativity controlled via optical pump power. Purple dots are the microwave-to-optical data. Orange dot is an optical-to-microwave measurement. Both data are expected to follow the black dashed theory line.
        }
        \label{fig4}
\end{figure}
The measurements in Fig. \ref{fig2} establish that this system achieves simultaneous, fully-resonant electro- and opto-mechanical couplings (with $C_\mathrm{om}>1$). This motivates us to evaluate the performance of this platform from the perspective of quantum transduction, quantifying the efficiency with which it converts microwave photons to optical photons.
In our system, the transduction efficiency $\eta$ (defined as the scattered optical photons exiting the Fabry-Perot cavity per microwave photon entering through the coupling pin, see Fig. \ref{fig1}a) can be analytically expressed in terms of the electro/opto-mechanical cooperativities as\cite{andrews_bidirectional_2014, han_cavity_2020}
\begin{equation}\label{eq_eta}
    \eta = \eta_{\mathrm{opt}}\eta_\mu \frac{4C_{\mathrm{om}}C_{\mathrm{em}}}{(1+C_{\mathrm{om}}+C_{\mathrm{em}})^2}.
\end{equation}
Here, $\eta_{\mathrm{opt}}$ and $\eta_\mu$ are the optical and microwave port-coupling efficiencies (i.e., the fraction of the total loss rate attributable to the input/output port).
Thus, one seeks to have greater-than-unity (and equal) cooperativities for both subsystems, as well as efficiently-coupled ports for transferring photons in and out.
In the hybrid quartz cavity demonstrated in Fig.  \ref{fig2}, we reach coupling efficiencies of $\eta_\mu=0.43$ and $\eta_\mathrm{opt}=0.53$.
Higher coupling efficiencies (i.e. over-coupled resonators) are possible by adjusting the microwave coupling pin and using optical mirrors with imbalanced reflectivities.
We note that the modes of Gaussian optical resonators easily achieve high fiber-coupling efficiency, which is a key challenge for low loss integration.

Repeating the measurements of Fig.  \ref{fig2}(d), we can characterize the microwave-to-optical scattering parameter and extract $\eta_\mathrm{oe}$ (optical photons out per microwave photon in). Alternatively, we can inject a sideband into the optical signal mode, and directly measure the microwave response, to characterize $\eta_\mathrm{eo}$ (microwave photons out per optical photon in). Both measurements are shown in Fig. \ref{fig4}(a), illustrating the bidirectionality of this device. The Fano dip in the $\eta_\mathrm{oe}$ data results from the previously-described electro-optic transduction.
The added noise floor of this transduction would be set by the thermal noise of the acoustic resonator, with an expected 16 phonons at 9K temperature. Characterization of the thermal noise in this quartz system was not carried out, but similar measurements on CaF\textsubscript{2} confirm good thermalization (see Sec. 5), consistent with past works\cite{kharel_multimode_2019}.
The higher noise floor in the $\eta_\mathrm{eo}$ data is a combination of Johnson noise of the room-temperature microwave detector, microwave amplifier noise, and detector noise.

On resonance, ($\Omega_\mathrm{m}/2\pi = 11.366$ GHz), we verify that the transduction efficiency scales as predicted with optical cooperativity.
The experimental data (purple points) in Fig.  \ref{fig4}(b) well-describes the theoretical curve (black dashed line) following Eq. \ref{eq_eta}, assuming $C_\mathrm{em} = 5.6\times10^{-8}$.
In particular, we confirm that the transduction is maximized for $C_\mathrm{om}=1$, implying that our optomechanical subsystem is sufficiently strongly coupled to saturate the transduction efficiency. Further increased $C_\mathrm{om}$ damps the acoustic resonance, reducing the overall efficiency.
Note that the model is well-matched to the data for $g_\mathrm{em}/2\pi$ = 347 Hz ($C_\mathrm{em} = 5.6\times10^{-8}$), instead of the simulated $g_\mathrm{em}/2\pi$ = 298 Hz ($C_\mathrm{em} = 4.18\times10^{-8}$).
This may be attributable to a discrepancy between the simulated microwave resonator and the assembled device.

The data presented in Fig. \ref{fig4} illustrates the current performance of this device, where we observe a maximum conversion of $\eta = 1.2^{+1.0}_{-0.6}\times 10^{-8}$, with a bandwidth of 500 kHz.
The uncertainty here comes from detector and cable loss calibrations.
With coupling efficiencies exceeding $10^{-1}$ and optomechanical cooperativity exceeding 1, the performance is currently limited by low electromechanical cooperativity. 
However, our ability to reach $C_\mathrm{om}=1$ indicates that we are maximally sensitive to forces acting on the resonator.
This motivates the possibility of using our platform as a sensitive phonon spectroscopy tool, with applications in materials science.


\section{Precision sensing of anomalous piezoelectricity}

\begin{figure}[h!]
    \centering
        \includegraphics[width=7.5cm]{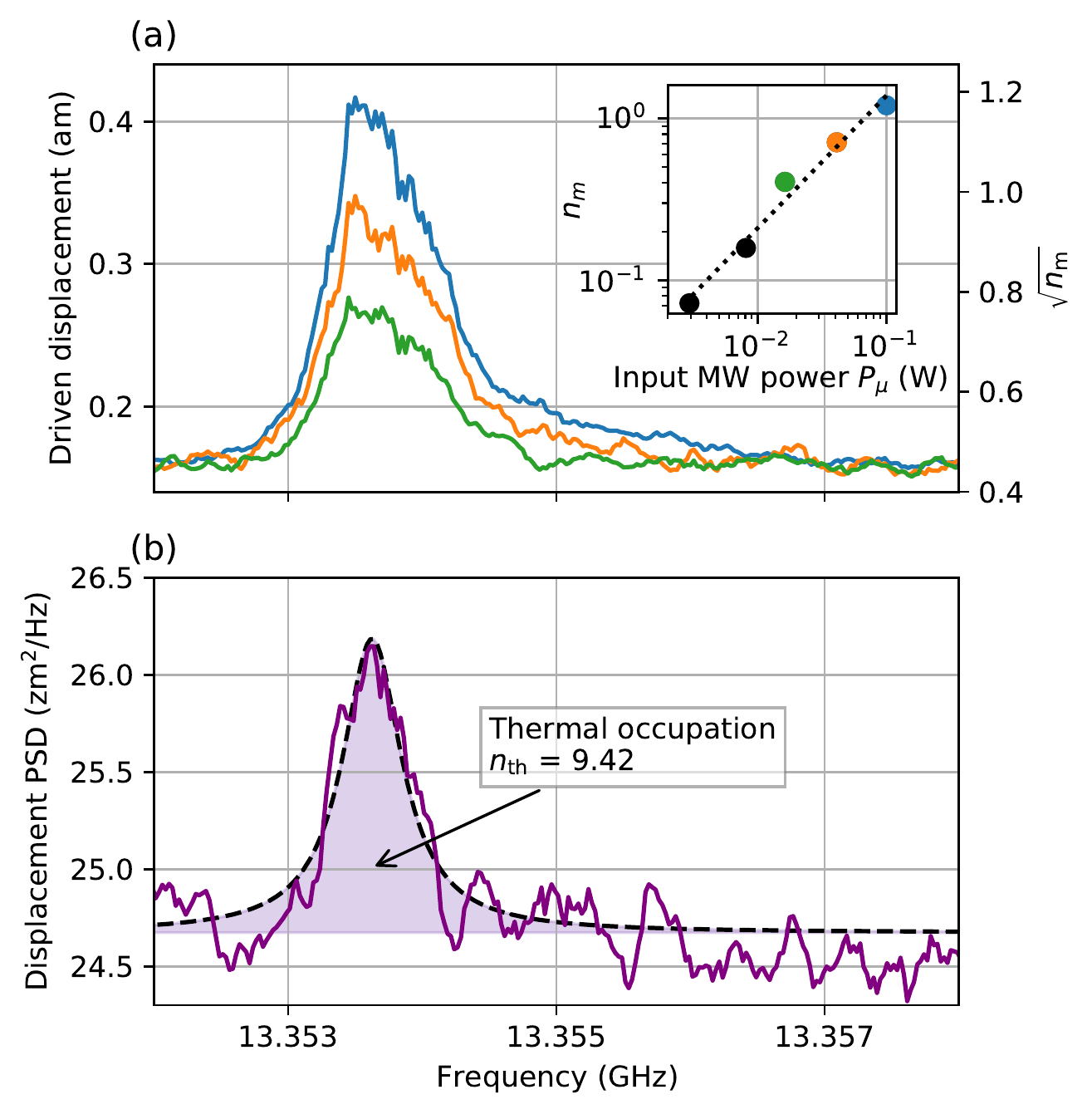}
        \caption{
        \textbf{Spectroscopy of anomalous piezoelectricity in CaF\textsubscript{2}}
        (a) Coherent, cavity-enhanced optical spectroscopy of microwave-driven motion in CaF\textsubscript{2}. The amplitude of driven motion (characterized by $\sqrt{n_\mathrm{m}}$ where $n_\mathrm{m}$ is phonon number) is calculated from the detected optical signal.
        The inset shows the linear relation (black dashed line) between the peak driven motion (in units of phonon number) and microwave power.
        To obtain the driven motion, background noise from the main figure is subtracted in the inset data.
        The color of each datapoint matches that of the main graph. Black points are not plotted in the main graph.
        (b) Amplitude spectral density of undriven (thermal) motion. PSD: power spectral density.
        Black dashed line is a Lorentzian fit with 535 kHz linewidth.
        Shaded area under the Lorentzian fit is integrated to obtain the effective mechanical occupation of $n_\mathrm{th}$ = 9.42.
        The peak of the Lorentzian fit indicates the detection sensitivity, corresponding to 2.1$\times10^{-4}$ phonons$\cdot\mathrm{Hz}^{-1}$
        }
        \label{fig3}
\end{figure}
This sensitivity, combined with the modular construction of the system, makes it possible to investigate electromechanical couplings in a variety of materials.
In particular, it enables us to probe possible anomalous piezoelectricity in non-piezoelectric materials.
There is strong motivation for this type of study, as parasitic piezoelectricity may be a relevant loss channel for emerging solid state quantum technologies \cite{ioffe_decoherence_2004, scigliuzzo_phononic_2020, diniz_intrinsic_2020}. 
Subsurface lattice damage, or even the inherent asymmetry of the lattice boundary\cite{georgescu_surface_2019}, can disrupt the centro-symmetry of nominally non-piezoelectric materials.

One such material not expected to have a piezoelectric response is CaF\textsubscript{2}.
The Brillouin-active mechanical frequency for CaF\textsubscript{2} is $\sim$13.3GHz, which we can easily target by adjusting the microwave resonator and optical FSR.
Doing so, we can repeat the experiment from Fig. \ref{fig2}(d), applying a microwave drive at the CaF\textsubscript{2} Brillouin frequency and looking for an optical response.
The results of this experiment are shown in Fig. \ref{fig3}(a), where we clearly see non-zero microwave-driven motion, indicating a measurable electro-mechanical coupling
(see Supplement for experimental parameters).
%
The detected signal only exists at the Brillouin frequency, indicating that it is indeed a mechanical response. This rules out the possibility that the response is originating from an unexpected electro-optical coupling.
%
As shown in the inset of Fig. \ref{fig3}(a), we observe a linear electromechanical coupling. We refer this coupling to anomalous piezoelectricity, although the innate mechanisms to such effect are uncertain (eg. charged surfaces and crystal defects).
%
Subtracting background noise (consisting of shot noise and detector noise) and thermal noise from the response shown in Fig.\ref{fig3}(a), we detect coherent acoustic response corresponding to $n_\mathrm{m}$=1.2, or a peak displacement for our acoustic standing wave of 0.25 am.
This corresponds to an effective electromechanical coupling rate of $g_{\mathrm{em}}^{\mathrm{CaF_2}}/2\pi$ = 0.03 Hz.

This effective piezoelectricity could be attributed to imperfections distributed in the bulk or concentrated at the surface.
If the acoustic response was derived from a piezoelectric coupling across a thin surface layer of 1 nm\cite{georgescu_surface_2019, ioffe_decoherence_2004}, it would correspond to an effective piezoelectric constant of $d_{33}$ = 2.44 pm/V (comparable to common piezoelectric materials).
However, such a large piezoelectric coefficient would be inconsistent with other investigations into CaF\textsubscript{2}\cite{aktas_piezoelectricity_2021}, and quantitative interferometric-based piezo force microscopy (IDS-PFM)\cite{labuda_quantitative_2015, collins_quantitative_2019} measurements conducted on this sample (which revealed no signal, see Supplement).
Alternatively, if it is derived from a uniform piezoelectric coupling distributed through the bulk of the crystal\cite{robinson_electrical-mechanical_1978, sharma_possibility_2007, aktas_piezoelectricity_2021}, it would correspond to a weak effective $d_{33}$ of 0.083 fm/V,
which would be consistent with the previously mentioned null results from IDS-PFM (given measurement noise floor of 55 fm/V).
To explain this effective bulk property, one might consider an ensemble of polar defects, which effectively act as piezoelectric emitters.
Such a mechanism could be intrinsically linked to dielectric loss, motivating further study, particularly in the context of solid state qubit decoherence.

As in transduction, the fundamental noise floor of this technique is ultimately limited by the thermal noise of the acoustic resonator, which we can detect in the absence of the microwave drive (see Fig. \ref{fig3}b).  
The effective thermal motion we detect corresponds to an average occupation of $n_\mathrm{th}$ = 9.42 (at $C_\mathrm{om} = 0.4$), consistent with thermalization to the cryostat ($T$ = 8.7K) plus some laser cooling by the pump.
Note that we observe no significant heating despite $>100$mW of input pump power, highlighting the good thermal anchoring and minimal absorption in this bulk platform.
On resonance, this thermal-limited noise floor is 162 zm (0.21 phonons) in the 1kHz bandwidth of our driven measurement, which corresponds to a piezoelectric coefficient sensitivity of 7 am/V, or 2.2 am/V at a bandwidth of 100 Hz in the case of an evenly distributed bulk piezoelectricity.
This compares favorably to existing techniques for probing piezoelectricity, such as resonant piezoelectric spectroscopy (RPS), resonant ultrasound spectroscopy (RUS) \cite{aktas_piezoelectricity_2021}, and PFM \cite{labuda_quantitative_2015, collins_quantitative_2019}. Critically, our technique extends this sensitivity to GHz frequencies and cryogenic temperatures. Applying this new materials analysis tool to materials like silicon, sapphire, and diamond can be of immediate relevance to quantum technologies\cite{georgescu_surface_2019, ioffe_decoherence_2004, scigliuzzo_phononic_2020, diniz_intrinsic_2020, maze_properties_2011}.

\section{Outlook and conclusion}

\begin{table}
    \centering
    \begin{tabular}{|c|c|} 
        \hline
         & $C_{\mathrm{em}}$  \\ \hline\hline
        Current experiment & $5.6\times10^{-8}$\\ \hline
        Stronger piezoelectric material
        & $\uparrow 50\times$\\ \hline
        Acoustic mode optimization & $\uparrow 80\times$\\ \hline
        Microwave mode optimization
        & $\uparrow 10\times$\\ \hline
        Higher Microwave cavity-Q
        & $\uparrow 10^2\times$\\ \hline
        Higher acoustic cavity-Q
        & $\uparrow 10^3\times$\\ \hline
        Combined improvements & $\uparrow \sim10^{9}\times$\\ \hline
    \end{tabular}
    \caption{\textbf{Electromechanical cooperativity improvements}
    System modifications to improve $C_\mathrm{em}$, contributing to both the transduction efficiency and piezo-detection sensitivity.
    }
    \label{table_improve}
\end{table}

Much progress has been made in both electro-optic and piezo-opto-mechanical transduction in recent years, reaching $\eta\sim \mathcal{O}(10^{-1})$ with minimal added noise\cite{sahu_quantum-enabled_2022, mirhosseini_superconducting_2020} and achieving preliminary integration with superconducting qubits\cite{mirhosseini_superconducting_2020, delaney_superconducting-qubit_2022}.
However, given the qualitative advantages of this platform, including robust thermal properties and power handling, high optical collection efficiency, and modularity, it is interesting to consider the outlook for efficient transduction.

Since $C_\mathrm{om}$, $\eta_\mathrm{opt}$, and $\eta_\mu$ are of order unity, future transduction improvements must target $C_\mathrm{em}$. First, since $C_\mathrm{em}\propto g_\mathrm{em}^2\propto d_{33}^2$, a stronger piezoelectric such as LiNbO\textsubscript{3} or BaTiO\textsubscript{3} can immediately improve $\eta$ (e.g. by 50$\times$ for LiNbO\textsubscript{3}).
Recalling that this platform requires no nanofabrication on the sample, it is straightforward to make such material changes.
There is also room for further optimization of mode-matching between the acoustic and microwave resonators. For example, a larger acoustic waist could increase $C_\mathrm{em}$ by 80$\times$, and a re-entrant microwave cavity (with increased field concentration) could increase $C_\mathrm{em}$ by $\sim$ 10$\times$.
Other extensions based on planar microwave resonators, or piezoelectric superlattices may offer additional gains in electromechanical coupling, and merit further investigation.

There is also significant room to improve the lifetimes of the component microwave and acoustic resonators, increasing the electromechanical cooperativity.
For instance, this demonstration uses a non-superconducting copper microwave cavity, with a modest $Q<1000$. 
Moving to a superconducting metal, there is ample evidence for both post and re-entrant cavities with Q $>10^7$ \cite{reagor_quantum_2016}.
While co-integrating superconducting resonators with optics is a known technical challenge \cite{mirhosseini_superconducting_2020}, this macroscopic, 3D platform may offer increased robustness of the superconducting resonator \cite{hease_bidirectional_2020}.
Additionally, while current acoustic performance is limited by diffraction loss, newly developed fabrication techniques allow us to produce stable plano-convex resonators with $Q>10^7$  \cite{kharel_ultra-high-q_2018}.
Implementation of acoustic resonator with better Q also allows us to reach $C_\mathrm{om}>1$ with weaker pump power (e.g. with $Q_\mathrm{m} = 10^7$, $C_\mathrm{om} = 1$ only requires $P_\mathrm{p}\approx 1\mu\mathrm{W}$), increasing compatibility with mK cryogenic systems.

Together, these improvements highlight a feasible path from the $\eta=1.2\times10^{-8}$ demonstrated here towards bidirectional transduction with near unity efficiency.  
Within the landscape of transduction platforms, piezo-Brillouin systems such as this have a unique set of advantages and constraints, and merit further investigation for quantum applications.

Beyond transduction, we have demonstrated how this system presents a versatile platform for materials analysis and other technological pursuits.
Specifically, Brillouin optical readout offers the ability to conduct precision spectroscopy of GHz motion with single-quanta sensitivity (i.e. a noise floor of 5.12 zm$\cdot \mathrm{Hz}^{-1/2}$). 
We have demonstrated how this can be used to investigate parasitic piezoelectricity, with detectable electromechanical coupling as low as $g_\mathrm{em}^{\mathrm{CaF_2}}/2\pi = 0.03$ Hz. 
Note that most of the improvements of Table 1 would also benefit such spectroscopy efforts, with the possibility of detecting parasitic couplings that are many orders of magnitude smaller.
Such measurements open the door to further evaluation of substrate purity and surface treatments, of critical importance in quantum technologies.
The dual microwave/optical functionality may also be useful for investigating strain-active solid state defects. Specifically, the resonant piezoelectric drive presents a mechanism for rapidly actuating strain fields, and the optical cavity offers the possibility of Purcell-enhanced photon collection.

\subsection{Acknowledgments}
We thank F. Ruesink, Y. Luo, S. Gertler, S. Ganjam, A. Read, N. Jin, Y. Zhou, M. Pavlovich, H. Cheng, and Y. Dahmani for helpful discussions.
Facilities use was supported by Yale SEAS cleanroom, Yale West Campus cleanroom, and Yale Gibbs machine shop.

This research was initially supported by the U.S. Department of Energy, Office of Science under award number DE-SC0019406 and completed under support by the U.S. Department of Energy, Office of Science, National Quantum Information Science Research Centers, Co-design Center for Quantum Advantage (C2QA) under contract number DE-SC0012704.

Piezoresponse force microscopy research was supported by the Center for Nanophase Materials Sciences (CNMS), which is a U.S. Department of Energy, Office of Science User Facility at Oak Ridge National Laboratory. This paper has been authored by UT-Battelle, LLC with the U.S. Department of Energy.

\subsection{Data availability} Data underlying the results presented in this paper are not publicly available at this time but may be obtained from the authors upon reasonable request.

\subsection{Supplemental document} See Supplement for supporting content.

\bibliographystyle{naturemag}

\bibliography{hybrid_v2.bib}

\end{document}


\maketitle

\tableofcontents
\clearpage

\section{Full apparatus setup}
\begin{figure}[htbp]
    \centering
    \includegraphics[width=\textwidth]{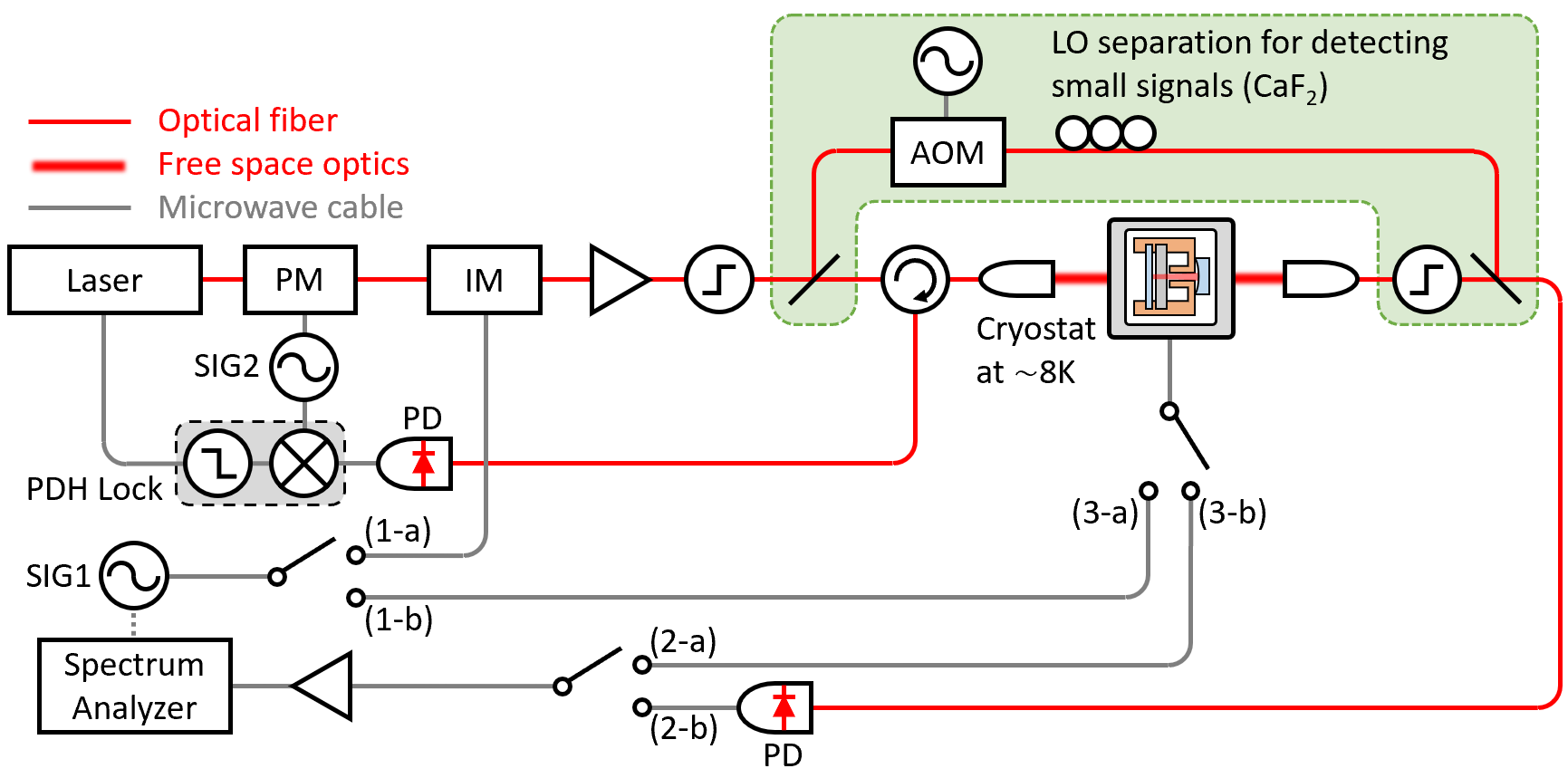}
    \caption{
    \textbf{Experimental apparatus}
    PM: phase modulator, IM: intensity modulator, AOM: acousto-optic modulator, PDH: Pound-Drever-Hall, SIG: signal generator.
    For OMIT measurement, switch combination (1-a) and (2-b) (3-a) are used (switch 3 is irrelevant).
    For microwave to optical conversion, switch combination (1-b), (2-b) and (3-a) are used.
    For optical to microwave conversion, switch combination (1-a), (2-a) and (3-b) are used.
    Part of the apparatus shaded in green is not used for X-cut quartz measurement, but instead only used for CaF\textsubscript{2} measurement to achieve better signal-to-noise ratio for highly sensitive piezoelectricity detection.
    }
    \label{fig:SI_fig1}
\end{figure}

The experimental setup is illustrated in Figure \ref{fig:SI_fig1}.
The laser is locked to the optical cavity via the Pound-Drever-Hall (PDH) locking technique. The optical sidebands required for the lock are generated by the signal generator (SIG2) and the phase modulator (PM).
For optomechanically-induced transparency (OMIT) measurements, a sideband is generated by an intensity modulator (IM), which is driven by a microwave signal generator (SIG1) synchronized to the readout frequency of a spectrum analyzer.
Here, the main optical tone and the sideband act as the pump and probe, respectively.
The beat note between the transmitted pump and probe is measured via a high speed photodetector and spectrum analyzer. The readout frequency of the spectrum analyzer can be synchronized to the frequency of the signal generator, effectively functioning as a scalar network analyzer.
In the microwave-to-optical transduction configuration, the drive tone from the signal generator directly drives the microwave cavity, piezoelectrically exciting phonons from which the pump light scatters to achieve electro-optical conversion.
For optical-to-microwave transduction, the optical probe is generated from the pump using the intensity modulator. However, instead of measuring the optical response of the probe (as in OMIT), the spectrum analyzer directly reads the signal leaving the microwave cavity.

For the OMIT/transduction measurements performed on quartz, the transmitted pump serves as a local oscillator (LO) for high-frequency heterodyne detection of the signal sideband. In this configuration, it is not easy to attenuate the LO power (e.g. to avoid detector saturation) without also attenuating the signal.
Therefore, for enhanced sensitivity in the CaF\textsubscript{2} measurements, we modify the circuit (green shaded box of Figure \ref{fig:SI_fig1}), filtering out the transmitted pump and using a separate, controllable LO derived from the original laser.
We also shift the frequency of this new LO using an Acousto-Optic Modulator (AOM) to avoid microwave crosstalk in the detection channel.
The spectrum analyzer frequency can be offset accordingly to accommodate this shift between drive frequency and heterodyne frequency, resulting in a unique signal frequency to avoid any potential microwave crosstalk. Using this separate-path LO can introduce low-frequency phase noise ($<$1kHz) in the heterodyne beat note, which does not impact our measurements.

\section{Individual cavity characterization}

\begin{figure}[t]
    \centering
    \includegraphics[width=\textwidth]{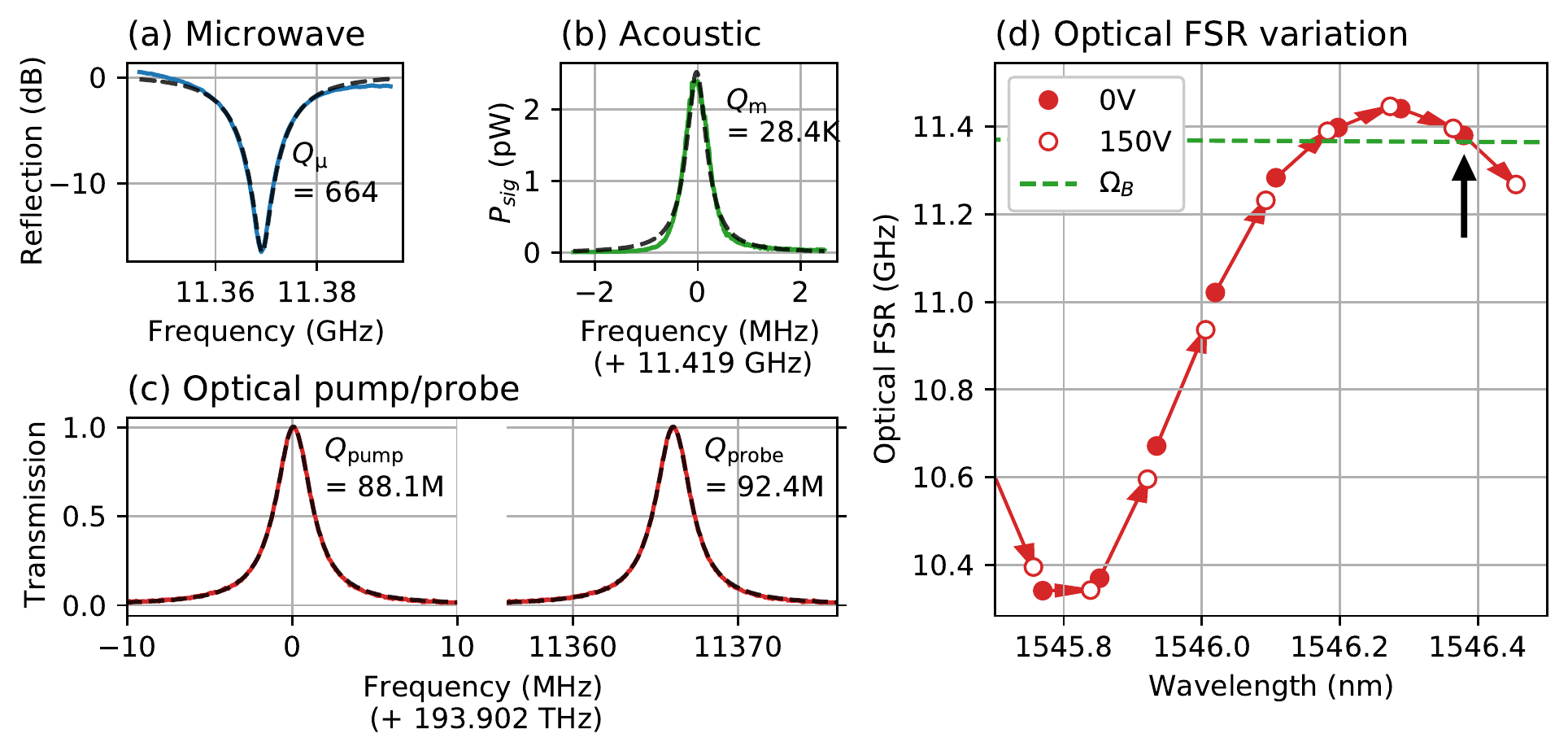}
    \caption{
    \textbf{Cavity loss and tunability characterization}
    (a) Microwave, (b) acoustic, and (c) optical cavity spectrum with corresponding Q-factor. (d) Illustration of optical cavity mode-spacing tunability. Red datapoints show varying FSR at different optical wavelength whereas red arrow shows the tuning range we could achieve with a piezo actuator controlling the optical cavity length. Black arrow indicates one of the crossing points between optical cavity mode-spacing and Brillouin frequency where we chose to operate at.
    }
    \label{fig:SI_fig2_cavity}
\end{figure}

In section 3 of the manuscript, we list the loss rates and the Q factors (finesse for optical cavity) of optical, acoustic, and microwave cavities ($\kappa_\mathrm{opt}/2\pi = 2.2 \mathrm{MHz}, \mathcal{F} = 5170, \Gamma/2\pi \approx 500 \mathrm{kHz}, Q_\mathrm{m} = 22800, \kappa_\mu/2\pi = 17.1 \mathrm{MHz}, Q_\mu = 664$) used for x-cut quartz demonstration of piezo-Brillouin spectroscopy.
Figure \ref{fig:SI_fig2_cavity}(a-c) are the cavity spectroscopy data with fits for x-cut quartz. 

Figure \ref{fig:SI_fig2_cavity}(a) shows a normalized reflection measurement of the microwave resonator, from which we find $\kappa_{\mu\mathrm{,i}}/2\pi = 9.79$ MHz $(Q_{\mu\mathrm{,i}} = 1160)$, $\kappa_{\mu\mathrm{,c}}/2\pi = 7.33$ MHz $(Q_{\mu\mathrm{,c}} = 1550)$, giving a total of $\kappa_\mu/2\pi = 17.1$ MHz $(Q_\mu = 662)$.
We use a coaxial stub cavity as the microwave cavity. The body of cavity is made out of OFHC copper, while a brass screw with a center through-hole is used as the center pin in order to achieve high tunability in resonance frequency. The seam loss between the copper body and the brass center pin results in a rather modest microwave Q factor in the current design.
Figure \ref{fig:SI_fig2_cavity}(c) shows (normalized) transmission measurements for the two optical cavity modes used as pump and signal mode.  We use optical sidebands to calibrate a linear frequency sweep over the modes, from which we can extract their linewidths.  A broader, calibrated wavelength scan is used to identify the mode spacing.  Due to the dielectric interface within the cavity, it is possible to have asymmetric cavity loss rates through the two mirrors, even though they have nominally identical reflectivities. Measurements of the resonant transmission and reflection are used to extract the relative loss rates through each mirror, as well as the internal loss. For both pump/probe modes of the optical cavity, we observe coupling rates of $\kappa_\mathrm{opt,c1}/2\pi \approx 0.7$ MHz and $\kappa_\mathrm{opt,c2}/2\pi \approx 1.2$ MHz, giving an internal loss rate of $\kappa_\mathrm{opt,i}/2\pi \approx 0.3$ MHz and combined loss rate of $\kappa_\mathrm{opt}/2\pi \approx 2.2$ MHz.
Figure \ref{fig:SI_fig2_cavity}(b) presents characterization data for an acoustic mode in x-cut quartz.  Specifically, we present single-pass piezo-Brillouin spectroscopy data (for which the optomechanical cooperativity is small, ensuring that we measure the bare mechanical linewidth in the absence of any optomechanical backaction).

\subsection*{Optical cavity FSR tunability}

The Fabry-Perot optical cavity used in our platform consists of two dielectric mirrors with 99.9\% reflectivity and a dielectric substrate (quartz or CaF\textsubscript{2}) in the middle.  The non-zero reflection of optical mode at the boundary of the dielectric media produces an intrinsic modulation of optical mode spacing. For the cavity geometry used here (11 mm vacuum, 0.5 mm quartz), we find modulation by $\pm$ 0.5 GHz around a mean spacing of 10.9 GHz. This variation allows us to find a mode spacing that approximately matches the Brillouin frequency. In the experiments here, we pump the lower-frequency mode, using the upper-frequency mode to resonantly match the anti-Stokes sideband.
Since the mode spacing variation is larger than the optical cavity linewidth, the other (Stokes) process is strongly suppressed.

\emph{In situ} fine-tuning of optical cavity mode spacing at cryogenic temperatures is achieved through the use of a piezo-actuator that allows translation of one mirror position. With this, we can match the optical mode spacing to the acoustic mode within the both optical and acoustic linewidths ($\Delta_\mathrm{opt} \approx \Omega_\mathrm{m}$).

The frequency tuning range is shown in Figure \ref{fig:SI_fig2_cavity}(d), where we clearly see greater than than 10\% ($>$ 1 GHz) intrinsic variation, plus fine-tuning via the piezo-actuator voltage (0V $\sim$ 150V) to match the optical mode spacing (red arrow) with the acoustic mode (green dashed line). In our experiment, we operate at optical wavelength of $\sim$ 1546.4 nm, indicated in back arrow in the figure.

\section{Coupling rate characterizations}

\subsection{Electromechanical coupling rate}

The electromechanical interaction Hamiltonian can be expressed as a product between piezo-induced stress field ($\sigma(\bar{r}) = \overline{\overline{c}} : \overline{\overline{d}}\cdot E(\bar{r})$) and strain field in the substrate ($s(\bar{r})$), given as $H_{\mathrm{int}}=\int{\sigma(\bar{r})s(\bar{r}) dV}$, where $\overline{\overline{c}}$ is the stiffness tensor, $\overline{\overline{d}}$ is the piezoelectric tensor, and $:$ is tensor product.
%
In our system, the coupling occurs predominantly between the z-directional electric and strain field due to the microwave cavity design and material properties.
Thus we consider longitudinal strain fields (${s}(\bar{r}) = {S}_z(\bar{r})$) interacting with an electric field along the z-axis (${\sigma}(\bar{r}) = c_{33}d_{33}{E}_z(\bar{r})$), resulting in
$H_{\mathrm{int}}=c_{33}d_{33} \int{{E}_z(\bar{r}){S}_z(\bar{r}) dV}$.
Quantizing the electric (${E}_z(\bar{r})=\bar{E}_0(\bar{r})({a} + {a}^\dagger)$) and strain fields (${S}_z(\bar{r})=\bar{S}_0(\bar{r})({b} + {b}^\dagger)$), the interaction Hamiltonian becomes

\begin{equation} \label{eq:int_hamil}
    \hbar g_{\mathrm{em}} = c_{33}d_{33}\int_{\mathrm{pz}}{\bar{E}_0(\bar{r})\bar{S}_0(\bar{r}) dV}
    ,
\end{equation}
where $\int_{\mathrm{pz}}dV$ is a volume integral across the piezoelectric substrate.

Acoustic strain field is assumed to be a standing wave in longitudinal direction and a Gaussian shape in transverse direction with mode waist proportional to the probing optical mode waist\cite{kharel_high-frequency_2019} ($r_\mathrm{m} = r_{\mathrm{opt}}/\sqrt{2}$, where $r_\mathrm{m}$ is the acoustic mode waist and $r_{\mathrm{opt}}$ is the optical mode waist), resulting in an expression
\begin{equation} \label{eq:strain}
    \bar{S}_0(\bar{r})=s_0e^{-\frac{r^2}{r_\mathrm{m}^2}}\sin(\frac{2\pi z}{\lambda_\mathrm{m}})\hat{z}
    .
\end{equation}
$\lambda_\mathrm{m}$ is the acoustic wavelength, $s_0 = \sqrt{\frac{2\hbar\Omega}{c_{33}L_\mathrm{m} A_\mathrm{m}}}$ is the zero point strain, $L_\mathrm{m}$ is the crystal thickness, and $A_\mathrm{m}$ is the acoustic mode area.

The microwave mode is concentrated around the central pin of a coaxial microwave cavity, resulting in a spherical mode shape with a radius of $\sim$3 mm. Meanwhile, mode waist (radius) of the cylindrical acoustic mode is $\sim$ 50 um, which is substantially smaller than that of the microwave mode. Moreover, the thickness of the crystalline substrate varies between 0.5 mm and 1 mm, which is also smaller compared to the microwave mode waist. As a result, the electric field parallel to the phonon propagation ($\hat{z}$-direction) within the microwave/acoustic mode overlap volume can be assumed to be uniform,
\begin{equation} \label{eq:efield}
    \bar{E}_0(\bar{r}) = E_0 \hat{z}
    ,
\end{equation}
where $E_0 = \sqrt{\frac{\hbar\Omega}{2\epsilon_0\epsilon_rV_\mu}}$
is the uniform zero-point electric field strength within the overlapping mode volume. We can obtain Electric field of the system through HFSS (3D high frequency simulation software). It is convenient to output the electric field values of the microwave mode when the E-field is at its maximum and H-field is is zero for a stored energy of $\hbar\Omega$. E-field value from the simulation ($\bar{E}_\mathrm{sim}(\bar{r})$) follows, $\hbar\Omega = \frac{\epsilon_0\epsilon_r}{2}\int dV |\bar{E}_\mathrm{sim}(\bar{r})|^2$. Within the microwave/acoustic mode overlap volume, we see uniform simulated electric field ($\bar{E}_\mathrm{sim}(\bar{r}) = E_\mathrm{sim} \hat{z}$). In terms of zero-point field, simulated field amplitude can be expressed as, $E_\mathrm{sim}= 2E_0$.

Combining equations (\ref{eq:int_hamil}), (\ref{eq:efield}) and (\ref{eq:strain}), we obtain the electromechanical coupling rate, 
\begin{equation} \label{eq:g_em}
    \begin{aligned}
    g_{\mathrm{em}} &= \frac{1}{\hbar}c_{33}d_{33}s_0 E_0 \int_{\mathrm{pz}}{e^{-\frac{r^2}{r_\mathrm{m}^2}} \sin(\frac{2\pi z}{\lambda_\mathrm{m}}) dV}
    \\
    &= d_{33} E_\mathrm{sim} \frac{\lambda_\mathrm{m}}{\pi} \sqrt{\frac{\Omega c_{33} A_\mathrm{m}}{2\hbar L_\mathrm{m}}} \sin^2\left(\frac{\pi t_\mathrm{pz}}{\lambda_\mathrm{m}}\right)
    ,
    \end{aligned}
\end{equation}
where $t_{\mathrm{pz}}$ is the piezoelectric substrate thickness.

There are three distinct cases of piezoelectric coupling that one can consider. Firstly, piezoelectricity can be distributed evenly throughout the entire bulk of the acoustic substrate. In this case, acoustic modes with an even longitudinal index ($t_{\mathrm{pz}} = n\frac{\lambda_\mathrm{m}}{2}$ with even $n$) will have zero coupling rate ($\sin^2\left(\frac{\pi t_\mathrm{pz}}{\lambda_\mathrm{m}}\right) = 0$), while the odd index modes ($t_{\mathrm{pz}} = n\frac{\lambda_\mathrm{m}}{2}$ with odd $n$) will have maximal coupling ($\sin^2\left(\frac{\pi t_\mathrm{pz}}{\lambda_\mathrm{m}}\right) = 1$). Secondly, substrates may have surface piezoelectricity on one side of the substrate. In this case, piezoelectricity is concentrated within a thin surface layer on one surface with thickness $t_{\mathrm{pz}}$ ($t_{\mathrm{pz}}\ll\lambda_\mathrm{m}$), while the bulk remains non-piezoelectric, resulting in a simplification of the sin term in the coupling rate expression as $\sin^2\left(\frac{\pi t_\mathrm{pz}}{\lambda_\mathrm{m}}\right) = (\frac{\pi t_{\mathrm{pz}}}{\lambda_\mathrm{m}})^2$. Lastly, there is the case for surface piezoelectricity on both sides of the substrate. Here, the modulation of zero and non-zero coupling rate for even and odd indexed acoustic modes appears as in the bulk piezoelectricity case, while the non-zero coupling rate is proportional to $t_{\mathrm{pz}}^2$ as in the single-sided surface piezoelectricity case. With these observations in mind, we examine our experimental results in Section 5 (on X-cut quartz) and Section 7 (on CaF\textsubscript{2}) of the supplement.

Another aspect of $g_\mathrm{em}$ is that it has a factor of $1/\sqrt{L_\mathrm{m}}$. Although we are integrating along the full length of the substrate, the sin term results in values equivalent to integrating along an effective piezoelectric thickness of $t_\mathrm{pz} < \lambda_\mathrm{m}/2$. Hence the factor, $1/\sqrt{L_\mathrm{m}}$, from strain normalization term remains throughout the expression of $g_\mathrm{em}$. Such dependency of $g_\mathrm{em}$ in substrate thickness suggests that using thinner substrate is advantageous in achieving larger $g_\mathrm{em}$, which we will later discuss the supplement.

\subsection{Optomechanical coupling rate}

Here, we derive the Brillouin optomechanical coupling rate, closely following prior studies\cite{renninger_bulk_2018,kharel_high-frequency_2019}.
Specifically, we consider phase-matched photoelastic coupling between optical Fabry Perot modes and longitudinal bulk acoustic resonances. The electric field profile of the $j$-th optical cavity mode is given by
\begin{equation}
{E}_{j}(r,z,t) = E_{j,0} e^{-r^2/r_{\mathrm{opt}}^2} \mathrm{sin}(k_j z) ({a}_j(t) + {a}^\dagger_j(t))
\end{equation}
where $E_{j,0}=\sqrt{\frac{2\hbar\omega_j}{\epsilon_0\epsilon_rA_{opt}L_{opt}}}$ is the zero-point field, $\omega_j$ is the $j$-th optical cavity mode frequency, $A_\mathrm{opt}$ ($L_\mathrm{opt}$) is optical mode area (optical cavity length), $\epsilon_0$ ($\epsilon_r$) is vacuum (relative) permittivity, $k_j$ is $j$-th optical wavevector, and ${a}_j$ is the anhillation operator for $j$-th optical mode.  Note that for simplicity, this model neglects modifications of the optical field due to the combined vacuum/dielectric composition of the cavity.  Full modelling of the optical mode profile, in the presence of this dielectric interface, is possible through use of a transfer matrix model \cite{kharel_high-frequency_2019}.
For the acoustic modes, we write the displacement of the $i$-th longitudinal mode as
\begin{equation}
    {U}_z(r,z,t) = U_{i,0} e^{-r^2/r_m^2} \mathrm{cos}(q_i z) ({b}_i(t) + {b}^\dagger_i(t))
    ,
\end{equation}
where $U_{i,0}=\sqrt{\frac{2\hbar}{\rho\Omega_\mathrm{m} A_\mathrm{m} L_\mathrm{m}}}$ is the zero-point displacement, $\rho$ is crystal density, $q_i$ is $i$-th acoustic wavevector, and ${b}_i$ is the anhillation operator for $i$-th acoustic mode.
Since we are using a plano-plano (as opposed to plano-convex) crystal for the acoustic resonator, our acoustic mode waist is not independently determined by the crystal geometry, but rather defined by the optical mode waist.  Specifically, $r_\mathrm{m}=r_\mathrm{opt}/\sqrt{2}$.  

The photoelastic interaction Hamiltonian is given by
\begin{equation}
    H_{\mathrm{int}} =\frac{1}{2}\epsilon_0\epsilon_r^2p_{13}\int\left(\frac{\partial {U}}{\partial z}\right){E}^2_{\mathrm{opt}}dV
    ,
\end{equation}
where $p_{13}$ is the relevant photoelastic constant.
In anticipation of the phase-matching requirement, we will consider intermodal coupling involving the $j$ and $j+1$ optical modes with wavevectors/frequencies which satisfy $k_{j+1}+k_j=q_i$ and $\Omega_m=\omega_{j+1}-\omega_j$.  Since $\omega_j\approx\omega_{j+1}$, this yields the Brillouin phase-matching condition from the main text ($q_i = 2n\omega_j/c$).  Plugging in the appropriate mode definitions and making the rotating wave approximation, the interaction Hamltonian is 
\begin{equation}\label{SIeq_Hint1}
\begin{aligned}
    H_{\mathrm{int}} &= \frac{1}{2}\epsilon_0\epsilon_r^2p_{13}\int dV q_i U_{i,0} E_{j+1,0} E_{j,0} e^{-\frac{r^2}{r_m^2}} e^{-\frac{r^2}{r_{\mathrm{opt}}^2}} e^{-\frac{r^2}{r_{\mathrm{opt}}^2}} \mathrm{sin}(q_i z) \mathrm{sin}(k_{j+1} z) \mathrm{sin}(k_j z) ({a}^\dagger_{j+1}{a}_j{b}_i + H.C.)
    ,
\end{aligned}
\end{equation}
Note that here we have assumed the crystal begins precisely at $z=0$ (i.e. coincident with the mirror).  In reality, the longitudinal component of this overlap integral will depend sensitively on the position of the acoustic standing wave within the optical standing wave.  As shown in \cite{kharel_high-frequency_2019}, the approach we take here offers a good estimate of the \emph{maximum} possible coupling rate.  Experimental values may differ by up to a factor of 2, based on this exact longitudinal overlap.  This approximation, as well as the choice to neglect optical field redistribution due to the dielectric interface, constitute the main sources of uncertainty in predicting the cavity optomechanical coupling rate. 

The interaction Hamiltonian can be be alternatively written in terms of the single-photon coupling rate, $g_{\mathrm{om,0}}$, as following,
\begin{equation}\label{SIeq_Hint2}
    H_{\mathrm{int}} = \hbar g_{\mathrm{om,0}}({a}^\dagger_{j+1}{a}_j{b}_i+H.C.)
    .
\end{equation}
Combining Equation \ref{SIeq_Hint1} and \ref{SIeq_Hint2}, we obtain the single-photon coupling rate in the presence of an optical cavity,
\begin{equation} \label{SIeq_gom0}
    g_{\mathrm{om,0}} = \frac{\omega_j^2n^3p_{13}}{8c}\sqrt{\frac{2\hbar}{\Omega_\mathrm{m}\rho A_\mathrm{m} L_\mathrm{m}}}\frac{L_\mathrm{m}}{L_\mathrm{opt}}
    .
\end{equation}
The x-cut quartz OMIT measurements from Figure 2 of the main text offer a direct measurement of the cavity-enhanced coupling rate.  Using cavity parameters to calculate the intracavity power, we can infer an experimental value of $g_{\mathrm{om,0}}=5.28$ Hz which closely agrees with the analytical value of $g_{\mathrm{om,0}}= 5.89$ Hz from Equation \ref{SIeq_gom0}. For all data presented, experimental value is used unless specified.

Optical cavity-enhanced $g_{\mathrm{om}}$ is obtained by linearizing around a strong pump ($H_{\mathrm{int}} = \hbar g_{\mathrm{om,0}}{a}^\dagger_{j+1}{a}_j{b}_i \approx \hbar \sqrt{N_p}g_{\mathrm{om,0}}{a}^\dagger_{j+1}{b}_i = \hbar g_{\mathrm{om}}{a}^\dagger_{j+1}{b}_i$), giving the expression, $g_{\mathrm{om}} = \sqrt{N_\mathrm{p}}g_{\mathrm{om,0}}$, where $N_\mathrm{p}$ is the inter-cavity pump photon number. In our case, the maximum cavity-enhanced $g_\mathrm{om}$ achievable in our system is $g_\mathrm{om}=$ 1.4 MHz from OMIT ($N_\mathrm{p}\sim7\times10^{10}$ for 112 mW of pump power)

\subsection{Optomechanical coupling rate and cooperativity in single-pass configuration}

In single-pass configuration without an optical cavity, we can consider an optical cavity with $L_\mathrm{opt}=L_\mathrm{m}$ where the boundaries are defined by the acoustic crystal surfaces.
This removes the filling-factor term ($\frac{L_\mathrm{m}}{L_\mathrm{opt}} = 1$) from Equation \ref{SIeq_gom0} in describing the single-pass single-photon optomechanical coupling rate, resulting in an expression,
$g_{\mathrm{om,0}}'= \frac{\omega_j^2n^3p_{13}}{8c}\sqrt{\frac{2\hbar}{\Omega_\mathrm{m}\rho A_\mathrm{m} L_\mathrm{m}}}$.

Free-space cooperativity of an optomechanical system with an acoustic cavity but without an optical cavity is defined in reference \cite{renninger_bulk_2018} as
\begin{equation}
    C_\mathrm{om,0}^\mathrm{sp} = \frac{P_\mathrm{p}}{\hbar \omega_\mathrm{p}}\frac{L_\mathrm{m}}{v_\mathrm{g,p}}\frac{L_\mathrm{m}}{v_\mathrm{g,s}}\frac{g_\mathrm{om,0}'^2}{\Gamma}
    ,
\end{equation}
where $P_\mathrm{p}$ is the pump power, $\omega_\mathrm{p}$ is the optical pump wavelength, $v_\mathrm{g,p}$ ($v_\mathrm{g,s}$) is the group velocity of the pump (signal) light and $\Gamma$ is the acoustic loss rate.
Introducing group delay ($\tau = L_\mathrm{m} n/c$), and inter cavity photon number ($N_\mathrm{p} = P_\mathrm{p}L_\mathrm{m}/(\hbar \omega_\mathrm{p} v_\mathrm{g,p})$), the single-pass cooperativity can be re-written in a simpler form, $C_\mathrm{om}^{\mathrm{sp}} = \frac{g_{\mathrm{om}}^2}{\Gamma\tau^{-1}}$.

\section{Optical cavity enhancement of signal}

Here, we consider the specific configuration in which one seeks to measure an optical sideband scattered from a (piezoelectrically-driven) phonon population. In particular, we analyze the optical power in the scattered sideband on both single-pass and optical cavity configuration, to illustrate the benefit of the optical cavity.

\subsection{Signal without optical cavity enhancement}

In the absence of an optical cavity, a strong pump can still scatter off phonons to generate a Brillouin sideband.  Following reference \cite{renninger_bulk_2018}, the optical power in this sideband can be written as
\begin{equation} \label{eq:Psig_fs}
    P_{\mathrm{sig}}' = \left(\frac{g_{\mathrm{om,0}}'L_\mathrm{m}}{v_\mathrm{o}}\right)^2 P_\mathrm{p} n_\mathrm{m}
    .
\end{equation}
\subsection{Signal with optical cavity enhancement}
In the case where we do consider an optical cavity, it becomes natural to describe the optomechanical interaction through coupled equations of motion for the pump and signal fields. In the rotating frame, this can be written as,
\begin{equation}
    \begin{aligned}
        \dot{{a}}_\mathrm{s} &= i\sqrt{N_\mathrm{p}}g_{\mathrm{om,0}}{b}-\frac{\kappa_{\mathrm{opt}}}{2}{a}_\mathrm{s}
        \\
        \dot{{a}}_\mathrm{p} &= -\frac{\kappa_{\mathrm{opt}}}{2}{a}_\mathrm{p}+\sqrt{\kappa_{\mathrm{opt},c}}a_\mathrm{in}
        ,
    \end{aligned}
\end{equation}
where $a_\mathrm{in}$ is the external pump drive field ($P_\mathrm{p}/\hbar\omega_\mathrm{p}=\langle a_\mathrm{in}^\dagger a_\mathrm{in}\rangle$), ${a}_\mathrm{s}$ (${a}_\mathrm{p}$) is annihilation operator for the signal (pump) mode, ${b}$ is annihilation operator for phonons, $\kappa_\mathrm{opt}$ ($\kappa_\mathrm{opt,c}$) is optical cavity loss (coupling) rate.
A more extensive Hamiltonian approach will be taken in the later state space model section of the supplement. Certain simplifications are made in this section ($\Delta_o = 0, g_\mathrm{eo} = 0$).

In steady state, we obtain
\begin{equation} \label{eq:EOM_steady}
    \begin{aligned}
        i\sqrt{N_\mathrm{p}}g_{\mathrm{om,0}}{b}=\frac{\kappa_{\mathrm{opt}}}{2}{a}_\mathrm{s}
        \\
        \frac{\kappa_{\mathrm{opt}}}{2}{a}_\mathrm{p}=\sqrt{\kappa_{\mathrm{opt},c}} a_\mathrm{in}
        .
    \end{aligned}
\end{equation}
Inter-cavity pump photon number can then be written as
\begin{equation} \label{eq:int_cav_photon}
    \left(\frac{\kappa_{\mathrm{opt}}}{2}\right)^2N_\mathrm{p}=\kappa_{\mathrm{opt},c}\langle a_\mathrm{in}^\dagger a_\mathrm{in}\rangle
    .
\end{equation}
From input-output formalism, Equation \ref{eq:EOM_steady}, and Equation \ref{eq:int_cav_photon}, outgoing scattered signal field ($a_\mathrm{out}$) is
\begin{equation}
\begin{aligned}
    a_\mathrm{out} &= \sqrt{\kappa_{\mathrm{opt},c}}{a}_\mathrm{s}
    \\
    &=\left(\frac{\sqrt{\kappa_{\mathrm{opt},c}}}{\kappa_{\mathrm{opt}}}\right)^2
    (4ig_{\mathrm{om,0}})
    \sqrt{\langle a_\mathrm{in}^\dagger a_\mathrm{in}\rangle}{b}
    ,
\end{aligned}
\end{equation}
and the corresponding scattered signal power is
\begin{equation} \label{eq:Psig_cav}
\begin{aligned}
    P_{\mathrm{sig}} = 16 \left(\frac{\kappa_\mathrm{opt,c}}{\kappa_{\mathrm{opt}}^2}\right)^2 g_{\mathrm{om,0}}^2 P_\mathrm{p} n_\mathrm{m}
    .
\end{aligned}
\end{equation}

\subsection{Resonant enhancement of signal}
Comparing Equation \ref{eq:Psig_fs} and Equation \ref{eq:Psig_cav}, we get a ratio of
\begin{equation} \label{eq:Psig_enhancement}
    \begin{aligned}
        \frac{P_{\mathrm{sig}}}{P_{\mathrm{sig}}'} &=
        \frac{16\left(\frac{\kappa_{\mathrm{opt},c}}{\kappa_{\mathrm{opt}}^2}\right)^2 g_{\mathrm{om,0}}^2 P_\mathrm{p} n_\mathrm{m}}{\left(\frac{g_{\mathrm{om,0}}'L_m}{v_\mathrm{o}}\right)^2 P_\mathrm{p} n_\mathrm{m}}
        \\
        &= \frac{16}{\pi^2}\eta_\mathrm{opt}^2 \mathcal{F}^2
        .
    \end{aligned}
\end{equation}
Assuming everything was kept constant, we expect 16$\mathcal{F}^2$ improvement in signal power by introducing an optical cavity with with a finesse, $\mathcal{F}$.

\section{Transduction Spectrum}
\subsection{X-cut quartz: transduction}
\begin{figure}[t]
    \centering
    \includegraphics[width=1.\textwidth]{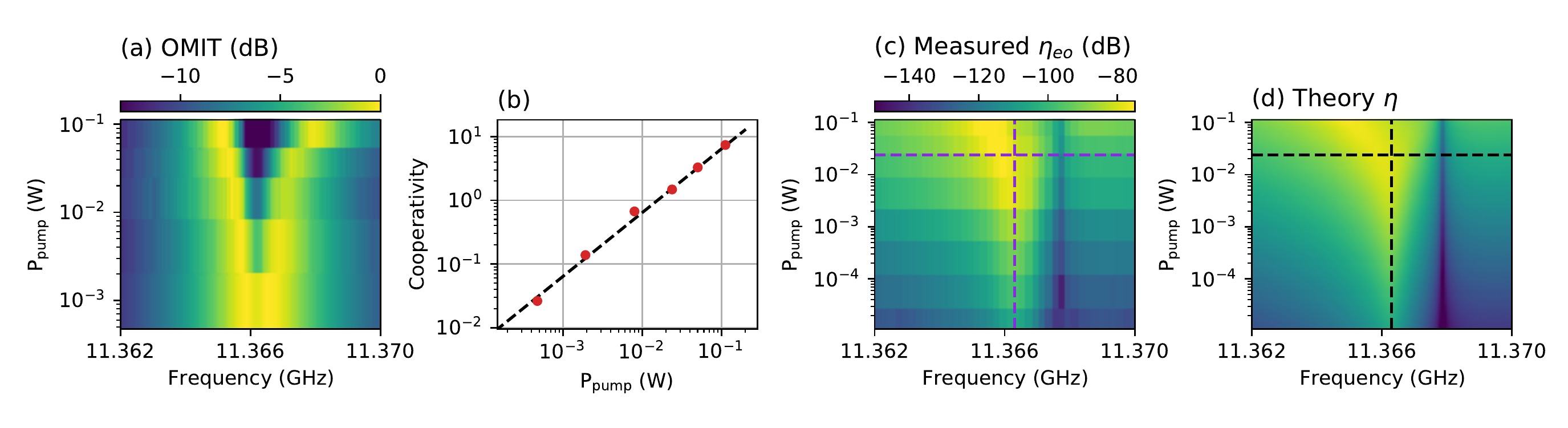}
    \caption{
    \textbf{Full OMIT and transduction spectrum for x-cut quartz}
    (a) colorplot of OMIT spectrum on x-cut quartz with varying optical pump power. Clear splitting of the optomechanical response indicates strong coupling. (b) Linear relation between cooperativity fitted from data in (a) and optical pump power. (c) Measured microwave-to-optical conversion at varying pump power. Shifting of transduction peak is clearly visible, indicating we are reaching $C_\mathrm{om}>1$. EO effect induced dip on the right side of the transduction peak is also visible. Dashed line in purple is the cross section along which Figure 3 in the main text is plotted. (d) Expected transduction spectrum closely following the measurement. Color scale is set to be identical to (c). Dashed line in black is the cross section along which Figure 3 in the main text is plotted.
    }
    \label{fig:SI_XQ_colorfig}
\end{figure}
Beyond what is presented in Figure 2 and Figure 3 in the main text, additional characterization can be made for x-cut quartz, as shown in Figure \ref{fig:SI_XQ_colorfig}. We performed a series of OMIT measurement at different optical pump powers (see Figure \ref{fig:SI_XQ_colorfig}a).
Here we observe normal mode splitting, which indicates optomechanical strong coupling ($2g_\mathrm{om}>\kappa_\mathrm{opt},\Gamma$), in addition to a large visibility in OMIT dip, indicating high optomechanical cooperativity ($C_\mathrm{om}>1$).
We can directly extract optomechanical cooperativity and confirm its linearity against optical pump power by fitting the OMIT data ($C_\mathrm{om}\propto N_\mathrm{p}\propto P_\mathrm{p}$, see Figure \ref{fig:SI_XQ_colorfig}b).
Microwave-to-optical transduction spectroscopy is measured across a range of optical pump power in Figure \ref{fig:SI_XQ_colorfig}(c).
The optical mode splitting from \ref{fig:SI_XQ_colorfig}(a) is also visible in the transduction spectra in \ref{fig:SI_XQ_colorfig}(c), particularly as the emergence of a left branch at high powers. The right branch is obscured due to interference with the electro-optic effect.
To accurately match this model to our data, note that we have to use $g_\mathrm{em}/2\pi = 347$ Hz (see Figure \ref{fig:SI_XQ_colorfig}d). This is close to the simulated coupling rate, $g_\mathrm{em}/2\pi = 298$ Hz, where the discrepancy can be attributed to the difference between simulated and machined microwave cavity and the shift in material properties in cryogenic conditions.
The theoretical model is based on a comprehensive state space model that describes piezoelectric EM, Brillouin OM, and Pockel EO interactions, laid out in the following section.

\subsection{State space model}
(This subsection closely follows the works of reference \cite{andrews_bidirectional_2014, han_cavity_2020}.) The Hamiltonian describing the system is
\begin{equation}
    H/\hbar = \omega_\mathrm{p}{a}_\mathrm{p}^\dagger{a}_\mathrm{p}+ \omega_\mathrm{s}{a}_\mathrm{s}^\dagger{a}_\mathrm{s}+ \Omega_\mathrm{m}{b}^\dagger{b}+ \Omega_\mu{c}^\dagger{c}+ (g_\mathrm{om,0}{a}_\mathrm{p}{a}_s^\dagger{b} + g_\mathrm{em}{b}^\dagger{c} + g_\mathrm{eo}{a}^\dagger{c} + \mathrm{H.c.})
    ,
\end{equation}
where $\omega_\mathrm{p}$ ($\omega_\mathrm{s}$) is optical wavelength of pump (signal) mode, $\Omega_\mu$ is microwave cavity resonance frequency, and ${c}$ is annihilation operator for microwave cavity mode. In the rotating frame with $H_0/\hbar = \omega_\mathrm{p}{a}_\mathrm{p}^\dagger{a}_\mathrm{p}+\omega_\mathrm{p}{a}_\mathrm{s}^\dagger{a}_\mathrm{s}$, we obtain an effective Hamiltonian
\begin{equation}
    H_{\mathrm{eff}}/\hbar = \Delta_\mathrm{o}{a}_\mathrm{s}^\dagger{a}_\mathrm{s}+ \Omega_\mathrm{m}{b}^\dagger{b}+ \Omega_\mu{c}^\dagger{c}+ (g_\mathrm{om,0}{a}_\mathrm{p}{a}_\mathrm{s}^\dagger{b} + g_\mathrm{em}{b}^\dagger{c} + g_\mathrm{eo}{a}^\dagger{c} + \mathrm{H.c.})
    ,
\end{equation}
where $\Delta_\mathrm{o} = \omega_\mathrm{s}-\omega_\mathrm{p}$ is the detuning of optical signal frequency from optical pump. In the undepleted pump regime, we can linearize the Hamiltonian about a strong, coherent pump, giving the expression
\begin{equation}
    H_{\mathrm{eff}}/\hbar = \Delta_\mathrm{o}{a}_\mathrm{s}^\dagger{a}_\mathrm{s}+ \Omega_\mathrm{m}{b}^\dagger{b}+ \Omega_\mu{c}^\dagger{c}+ (g_\mathrm{om}{a}_\mathrm{s}^\dagger{b} + g_\mathrm{em}{b}^\dagger{c} + g_\mathrm{eo}{a}^\dagger{c} + \mathrm{H.c.})
    ,
\end{equation}
where $g_\mathrm{om} = \sqrt{N_\mathrm{p}}g_\mathrm{om,0}$, and $N_\mathrm{p}$ is the intracavity pump photon number. The Heisenberg equations of motion for this Hamiltonian and the input-output relation between the fields are given by the following equations:
\begin{equation}
    \dot{\textbf{a}}(t) = A\textbf{a}(t) + B\textbf{a}_\mathrm{in}(t)
\end{equation}
\begin{equation}
    \textbf{a}_\mathrm{out}(t) = B^T\textbf{a}(t) - \textbf{a}_\mathrm{in}(t)
\end{equation}
\begin{equation}
    \textbf{a} = ({a}, {c}, {b})^T
\end{equation}
\begin{equation}
    \textbf{a}_\mathrm{in} = (a_\mathrm{in}, c_\mathrm{in})^T
\end{equation}
\begin{equation}
    \textbf{a}_\mathrm{out} = (a_\mathrm{out}, c_\mathrm{out})^T
\end{equation}
\begin{equation}
    A =
    \begin{pmatrix}
    -i\Delta_\mathrm{o}-\frac{\kappa_\mathrm{o}}{2} & ig_\mathrm{eo} & ig_\mathrm{om}\\
    ig_\mathrm{eo} & -i\Omega_\mu-\frac{\kappa_\mu}{2} & ig_\mathrm{em}\\
    ig_\mathrm{om} & ig_\mathrm{em} & -i\Omega_\mathrm{m}-\frac{\kappa_\mathrm{m}}{2}
    \end{pmatrix}
\end{equation}
\begin{equation}
    B =
    \begin{pmatrix}
    \sqrt{\kappa_\mathrm{o,c}} & 0\\
    0 & \sqrt{\kappa_{\mu\mathrm{,c}}}\\
    0 & 0
    \end{pmatrix}
    ,
\end{equation}
where \textbf{a} is a vector for resonator mode operators and \textbf{a}\textsubscript{in(out)} is a vector of input (output) fields.
This is a simplified state space model in which we only include ports that we care about -- in/output coupling ports. There are additional internal decay channels in this process that are not included in our model as we are assuming zero noise input.

In the frequency domain, we can reduce the expressions into a scattering matrix,
\begin{equation}
    \textbf{a}_\mathrm{out}(\omega) = S(\omega)\textbf{a}_\mathrm{in}(\omega)
\end{equation}
\begin{equation}
    S(\omega) = B^T(-i\omega I-A)^{-1}B-I
    .
\end{equation}
Using this scattering matrix, we can obtain various features that we can expect from our system, including OMIT, microwave-to-optical transduction, and optical-to-microwave transduction. Here we mainly focus on the transduction spectrum ($\eta(\omega) = |S_\mathrm{oe}(\omega)|^2 = |S_\mathrm{eo}(\omega)|^2$).
\begin{equation}
    \eta(\omega) = \frac{\kappa_\mathrm{o,c}}{\kappa_\mathrm{o}} \frac{\kappa_{\mu\mathrm{,c}}}{\kappa_\mu}
    \left|\frac{\alpha(\omega)}
    {\beta(\omega)}\right|^2
\end{equation}
\begin{equation}
    \alpha(\omega) = 2(-\sqrt{C_\mathrm{em}C_\mathrm{om}}+iC_\mathrm{eo}(1-\frac{i(\omega - \Omega_\mathrm{m})}{\Gamma/2}))
\end{equation}
\begin{equation}\label{eq_fulleta}
\begin{aligned}
    \beta(\omega) &= 2i\sqrt{C_\mathrm{em}C_\mathrm{om}C_\mathrm{eo}}+C_\mathrm{em}(1-\frac{i(\omega - \Delta_\mathrm{o})}{\kappa_\mathrm{o}/2})+C_\mathrm{om}(1-\frac{i(\omega - \Omega_\mu)}{\kappa_\mu/2})+C_\mathrm{eo}(1-\frac{i(\omega - \Omega_\mathrm{m})}{\Gamma/2})
    \\& + (1-\frac{i(\omega - \Delta_\mathrm{o})}{\kappa_\mathrm{o}/2})(1-\frac{i(\omega - \Omega_\mu)}{\kappa_\mu/2})(1-\frac{i(\omega - \Omega_\mathrm{m})}{\Gamma/2})
    .
\end{aligned}
\end{equation}
We observe peak conversion when the optical detuning, microwave mode, and the acoustic mode are in resonance. Measuring transduction at this resonant frequency ($\omega = \Delta_\mathrm{o} = \Omega_\mu = \Omega_\mathrm{m}$), we get
\begin{equation}\label{eq_fulletamax}
    \eta =
    \frac{\kappa_\mathrm{o,c}}{\kappa_\mathrm{o}} \frac{\kappa_{\mu\mathrm{,c}}}{\kappa_\mu}
    \frac
    {4(C_\mathrm{em}C_\mathrm{om}+C_\mathrm{eo}^2)}
    {4C_\mathrm{em}C_\mathrm{om}C_\mathrm{eo} + (C_\mathrm{em} + C_\mathrm{om} + C_\mathrm{eo} + 1)^2}
    .
\end{equation}
In the case of no electro-optic coupling, Equation \ref{eq_fulletamax} can be reduced into the expression for a typical electro-optomechanical transduction.
\begin{equation}
    \eta =
    \frac{\kappa_\mathrm{o,c}}{\kappa_\mathrm{o}} \frac{\kappa_{\mu\mathrm{,c}}}{\kappa_\mu}
    \frac
    {4C_\mathrm{em}C_\mathrm{om}}
    {(C_\mathrm{em} + C_\mathrm{om} + 1)^2}
\end{equation}

\subsection{Presence of electro-optic (EO) effect in transduction spectrum}

The expression of electro-optic (EO) coupling (Pockels effect) in materials is studied in various sources \cite{ilchenko_whispering-gallery-mode_2003, rueda_efficient_2016, holzgrafe_cavity_2020, mckenna_cryogenic_2020}. The single-photon electro-optic coupling rate between optical pump, optical signal, and microwave can be written as:
\begin{equation}
\begin{aligned}
    \hbar g_\mathrm{eo,0} &= \epsilon_0 \int_\mathrm{eo}{dV \epsilon_1 E_\mathrm{p} \epsilon_1 E_\mathrm{s} r_{13} E_\mu}
    \\
    &= \epsilon_0 \epsilon_1^2 r_{13} E_\mathrm{p,0} E_\mathrm{s,0}  E_{\mu\mathrm{,0}}\int_\mathrm{eo}{dV e^{-\frac{r^2}{r_{\mathrm{opt}}^2}} e^{-\frac{r^2}{r_{\mathrm{opt}}^2}} \mathrm{sin}(k_\mathrm{p} z) \mathrm{sin}(k_\mathrm{s} z)}
    ,
\end{aligned}
\end{equation}
where $r_{13}$ is the relevant linear electro-optic coefficient component. Electric fields of optical pump ($E_\mathrm{p}$), optical signal ($E_\mathrm{s}$) and microwave ($E_\mu$) is defined as in the previous sections, with corresponding normalization factors of $E_\mathrm{p,0} = \sqrt{2\hbar\omega_\mathrm{p}/\epsilon_0 \epsilon_1 A_\mathrm{opt}L_\mathrm{opt}}$ and $E_\mathrm{s,0} = \sqrt{2\hbar\omega_\mathrm{s}/\epsilon_0 \epsilon_1 A_\mathrm{opt}L_\mathrm{opt}}$. $E_{\mu\mathrm{,0}}$ is the zero-point electric field obtained through HFSS ($E_{\mu\mathrm{,0}} = E_\mathrm{sim}/2$), uniform along the interaction region.
Assuming identical optical pump and signal mode profiles in both transverse and longitudinal direction, the above expression simplifies into,
\begin{equation}
\begin{aligned}
    g_\mathrm{eo,0}
    &= \frac{1}{4} \epsilon_1 r_{13} E_\mathrm{sim} \omega_\mathrm{p} \frac{L_\mathrm{m}}{L_\mathrm{opt}}
    .
\end{aligned}
\end{equation}
EO coefficient in quartz is $r_{13}$ = 0.45 pm/V from crystal symmetry \cite{ivanov_direct_2018}.
Resulting single photon EO coupling rate in quartz is $g_\mathrm{eo,0}/2\pi$ = 1.05 mHz. Inside an optical cavity, this coupling rate is parametrically enhanced by the intercavity photon number ($g_\mathrm{eo} = \sqrt{N_\mathrm{p}}g_\mathrm{eo,0}$), reaching $g_\mathrm{eo}/2\pi$ = 137Hz.

\begin{figure}[t]
    \centering
    \includegraphics[width=.8\textwidth]{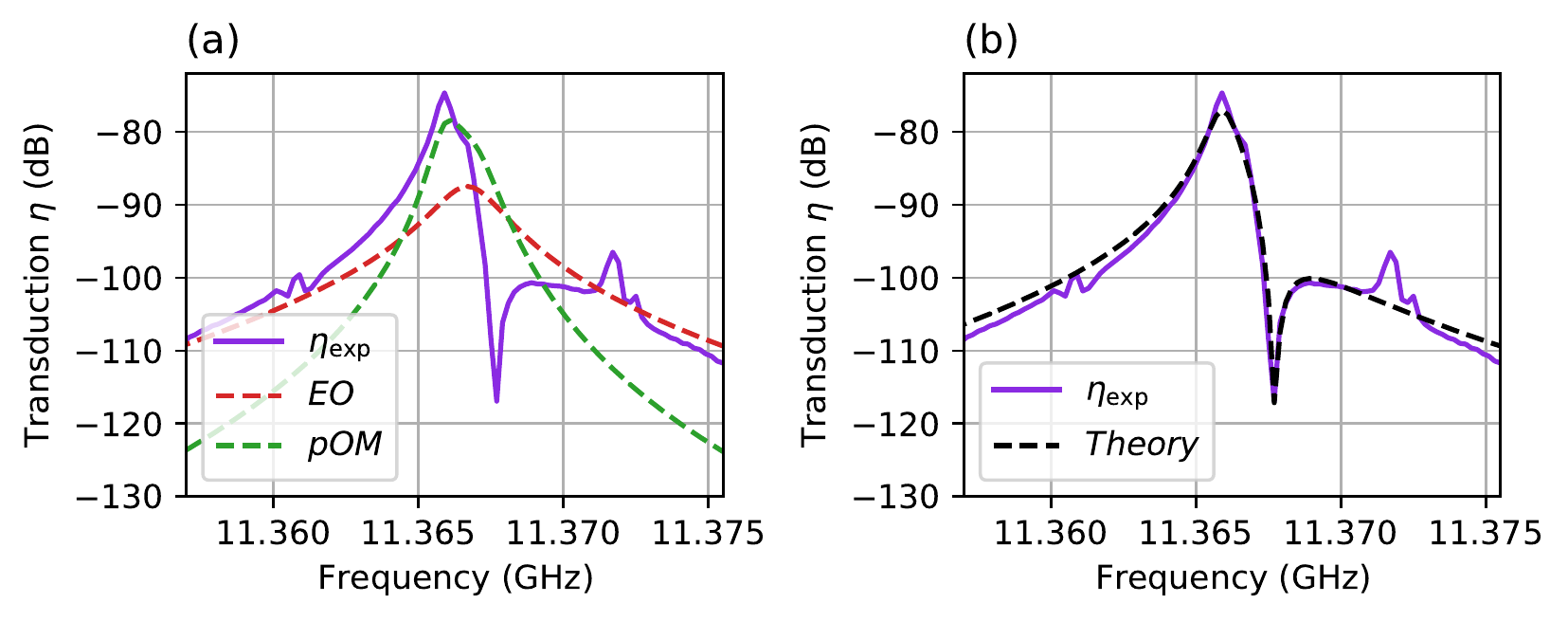}
    \caption{
    Microwave to optical conversion measured (purple) is fitted using state space model. The data is taken at $P_\mathrm{p}=$ 23.8 mW. Separate OMIT measurement indicates $g_\mathrm{om}/2\pi$ = 643 kHz and $C_\mathrm{om} = 1.48$. (a) Piezo-optomechanical (pOM) component (dashed green) and EO component (dashed red) of the fitted model is separately plotted against the conversion data. Fitted values of the coupling rates are $g_\mathrm{em}/2\pi$ = 347 Hz (theory value 298 Hz) and $g_\mathrm{eo}/2\pi$ = 162 Hz (theory value 137 Hz). (b) EO and pOM combined model (dashed black) fitted to data.
    }
    \label{fig:SI_fig3_EO}
\end{figure}

Incorporating a single-photon EO coupling rate into the state space model, we can compare our theoretical model to the transduction spectrum.
In fact, we can even observe individual contributions from piezoelectric coupling and EO coupling to the net transduction.
In Figure \ref{fig:SI_fig3_EO}, transduction from piezoelectricity is in dashed green and that from EO is in dashed red. The combined net transduction is in dashed black can be fitted to the experimental data in purple, showing good agreement.
Optomechanical coupling rate used in the fit is obtained through OMIT measurement ($g_\mathrm{om}/2\pi$ = 643 kHz and $C_\mathrm{om} = 1.48$ at $P_\mathrm{p} = 23.8$ mW).
The fitted parameters are $g_\mathrm{em}/2\pi$ = 347 Hz (theory value is 298 Hz) and $g_\mathrm{eo}/2\pi$ = 162 Hz (theory value is 137 Hz). Such difference can be attributed to a discrepancy between the simulated microwave cavity and the assembled device, and to the variation in material parameters at cryogenic temperatures.

The apparent interference between piezoelectric and EO effect near the center peak is because of similar EM and EO cooperativities. In the relatively lossy acoustic cavity regime that we are operating, we can still reach $C_\mathrm{om} > 1$. However, we have relatively low $C_\mathrm{em} = 5.6\times10^{-8}$, which is greater than $C_\mathrm{eo} = 2.8\times10^{-9}$ only by an order of magnitude. This is also observed in Figure \ref{fig:SI_fig3_EO}, where the peak of theoretical EO transduction and that of theoretical pOM transduction are separated by 10 dB. As a result, we observe a fano interference in the transduction spectrum where the pOM response crosses through EO response. This destructive interference is visible only on one side due to the phase relation between the pOM and EO interactions - constructive interference occurs on the other side.

One of the future paths to improve transduction, suggested in the manuscript, is using an acoustic cavity with better Q factor. Integration of a high-Q HBAR will bring about a significant increase in $C_\mathrm{em}$, while keeping $C_\mathrm{eo}$ constant, thus suppressing the formation of such fano-like feature in future implementations of this design.

\section{Potential improvements in transduction}

\begin{table}[t]
    \centering
    \begin{tabular}{|c|c|c|c|c|} 
        \hline
        Suggested improvements& $g_{\mathrm{em}}/2\pi$ & $C_{\mathrm{em}}$ & $\eta_\mathrm{opt}$($\eta_\mu$) & $\eta$\\ \hline\hline
        Current experiment & 347 Hz & $5.6\times10^{-8}$ & 0.53 (0.43) & $1.2\times10^{-8}$\\ \hline
        Optimized coupling efficiencies & - & - & $\sim$1 ($\sim$1) & $\uparrow 4 \times$\\ \hline
        X-cut quartz $\rightarrow$ Z-cut LiNbO3 & $\uparrow 7.3\times$ & $\uparrow 50\times$ & - & $\uparrow 50\times$\\ \hline
        Optimized acoustic mode geometry & $\uparrow 9\times$ & $\uparrow 80\times$ & - & $\uparrow 80\times$\\ \hline
        Re-entrant cavity & $\uparrow 3\times$ & $\uparrow 10\times$ & - & $\uparrow 10\times$\\ \hline
        Superconducting cavity & - & $\uparrow 10^2\times$ & - & $\uparrow 10^2\times$\\ \hline
        Plano-convex hBAR & - & $\uparrow 10^3\times$ & - & $\uparrow 10^3\times$\\ \hline\hline
        Combined improvement factors & $\uparrow \sim 200 \times$ & $\uparrow \sim 4 \cdot 10^9 \times$ & $\uparrow2\times$ ($\uparrow2\times$) & $\uparrow \sim 1.6 \cdot 10^{10} \times$ \\ \hline
        Combined improvement values & $\sim$68 kHz & $\sim$224 & $\sim$1 ($\sim$1) & $\sim$0.9 ($C_\mathrm{om} = C_\mathrm{em} = 10$) \\ \hline
    \end{tabular}
    \caption{\textbf{Electromechanical cooperativity improvements}
    }
    \label{table_improve}
\end{table}
Recall the transduction efficiency expression
\begin{equation}\label{eq_eta}
    \eta = \eta_{\mathrm{opt}}\eta_\mu \frac{4C_{\mathrm{om}}C_{\mathrm{em}}}{(1+C_{\mathrm{om}}+C_{\mathrm{em}})^2}
    .
\end{equation}
In our system, we can readily reach $C_\mathrm{om}$ of 1 and $\eta_\mathrm{opt}$ ($\eta_\mu$) of 0.53 (0.43).
Unity coupling efficiencies can be reached with ease by adjusting the microwave coupling pin to be over-coupled to the microwave cavity, and by having asymmetric optical cavity mirrors (i.e. creating a single-sided cavity).
Also note that the Gaussian optical cavity mode is well-suited for achieving high fiber-coupling efficiency, a key challenge for low loss integration of a transducer.
The expression therefore simplifies into
\begin{equation}\label{eq_eta}
\begin{aligned}
    \eta &\approx C_{\mathrm{em}}
    \\&=\frac{4g_\mathrm{em}^2}{\kappa_\mu \Gamma}
    ,
\end{aligned}
\end{equation}
in the limit of small $C_\mathrm{em}$ ($C_\mathrm{em} \ll C_\mathrm{om}$) and assuming unity coupling efficiencies ($\eta_\mathrm{opt}\approx\eta_\mu\approx 1$). Thus, improving the transduction efficiency boils down to improving $C_\mathrm{em}$ up to and above unity.

Among the approaches that we can take to enhance $C_\mathrm{em}$, one obvious choice is to achieve stronger $g_\mathrm{em}$ as it is quadratically related to $C_\mathrm{em}$. Rearranging Equation \ref{eq:g_em} in the bulk piezoelectric limit,
\begin{equation}
    g_{\mathrm{em}} = E_0 d_{33} c_{33} \sqrt{\frac{2A_\mathrm{m}}{\hbar \Omega_\mathrm{m} \rho L_\mathrm{m}}}
    .
\end{equation}
As laid out in Table \ref{table_improve}, one way we can improve $g_\mathrm{em}$ is through choosing a stronger piezoelectric material, such as LiNbO\textsubscript{3} or BaTiO\textsubscript{3}, instead of quartz.
In the case of Z-cut LiNbO\textsubscript{3}, piezoelectric tensor component increases from $d_{33} = 2.3$ pm/V to $d_{33} = 16.2$ pm/V, and stiffness tensor component from $c_{33} = 86.6$ GPa to $c_{33} = 244$ GPa.
Even taking into account the impact from the change in density ($\rho$ = 2650 kg/m\textsuperscript{3} to $\rho$ = 4630 kg/m\textsuperscript{3}), Brillouin frequency ($\Omega_\mathrm{m}$ = 11.4 GHz to $\Omega_\mathrm{m}$ = 21.0 GHz), and $E_0$ ($\frac{0.80}{\sqrt{2}}\times10^{-3}$ V/m to $\frac{0.53}{\sqrt{2}}\times10^{-3}$ V/m) due to higher relative permittivity, we can expect a higher $g_\mathrm{em}$ ($C_\mathrm{em}$) by 7.3$\times$ (50$\times$).

Another factor that influences $g_\mathrm{em}$ is the acoustic mode geometry. Increasing the acoustic mode waist allows us to achiever better mode matching between acoustic and microwave modes, while thinner substrate thickness allows for lower mass acoustic mode (with larger zero point motion) while maintaining same acoustic-microwave mode overlap.
Reasonable modifications can be made to the acoustic waist from 50 um to 200 um
and to the substrate thickness ($L_\mathrm{m}$) from 500 um to 100 um. This result in a higher $g_\mathrm{em}$ ($C_\mathrm{em}$) by 9$\times$ (80$\times$).
Note that reducing the substrate thickness will decrease the $g_\mathrm{om}$, yet it will not deter our ability to exceed unity optomechanical cooperativity ($C_\mathrm{om} > 1$) as the current setup reaches $C_\mathrm{om} \sim 10$ with relative ease.

Similarly, we can also improve the microwave and acoustic mode-matching by making modifications to the microwave cavity.
In doing so, we will be optimizing $E_0$, the remaining term in $g_\mathrm{em}$.
In order to reduce the microwave mode volume and concentrate the electric field, we can explore using a smaller coaxial microwave cavity or different cavity geometries such as a re-entrant cavity and other 2-D designs.
Simulating a re-entrant cavity, we obtain approximately $3\times$ stronger electric field compared to the coaxial design ($\frac{0.53}{\sqrt{2}}\times10^{-3}$ V/m $\rightarrow\frac{1.8}{\sqrt{2}}\times10^{-3}$ V/m), resulting in a higher $g_\mathrm{em}$ ($C_\mathrm{em}$) by 3$\times$ (10$\times$).

Besides optimizing $g_\mathrm{em}$, we can also improve on other factors in $C_\mathrm{em}$; namely, $\kappa_\mu$ and $\Gamma$.
This demonstration uses a non-superconducting cooper microwave cavity, with a modest Q $< 10^3$.
With superconducting materials, both post and re-entrant cavities can readily reach Q $> 10^8$.
Hence, in an ideal circumstance, $\kappa_\mu$ can be boosted by $> 10^5\times$.
However, we have to consider practical limitations such as the combination of high laser powers with superconducting resonators and the restriction in the bandwidth of the device that high Q microwave cavity may impose.
In reality, we will consider a conservative enhancement in $\kappa_\mu$ by $> 10^2\times$ as shown in Table \ref{table_improve}.
Diffractive loss in the acoustic cavity can be mitigated via shaping a concave surface on one side of the cavity through reactive-ion-etching.
Doing so will allow the acoustic cavity to form stable resonances with Q up to $\sim2.8\times10^7$\cite{kharel_ultra-high-q_2018} from the current $2.3\times 10^4$, allowing us a $\times 10^3$ improvement in $\Gamma$ at a cost in transduction bandwidth.

\section{Understanding CaF\textsubscript{2} data}

Experimental parameters for CaF\textsubscript{2} are as following: detector resolution bandwidth 1 kHz,
$\Omega_\mathrm{m}=$ 13.354 GHz,
$\Gamma = $ 535 kHz,
$L_\mathrm{m}=$ 0.5mm,
$A_\mathrm{m}= 2\pi\times(64\mathrm{um})^2$,
$\kappa_\mathrm{o}=$ 2.1 MHz,
$\kappa_\mathrm{o,c}=$ 0.598 MHz,
$E_0=5.1\times 10^{-4}$ V/m,
$\kappa_\mu=$ 22.4 MHz,
$\kappa_{\mu\mathrm{,c}}=$ 10.9 MHz.
For microwave driven motion measurement:
$P_\mathrm{p}=$ 23 mW,
$g_\mathrm{om}=$ 385 kHz,
$C_\mathrm{om}=$ 0.705.
For thermal phonon measurement:
$P_\mathrm{p}=$ 30.1 mW,
$g_\mathrm{om}=$ 427 kHz,
$C_\mathrm{om}=$ 0.914.

Conversion from detected signal power to phonon number is done following the expression, $P_{\mathrm{sig}} =  (\hbar \omega_\mathrm{p}) (\frac{4g_\mathrm{om}^2}{\kappa_\mathrm{o}\Gamma})\Gamma n_\mathrm{m}$, where $n_\mathrm{m}$ is the number of driven phonons.
In order to obtain piezoelectric coupling rate from the number of piezoelectrically driven phonons, we revisit the state space model for the equation of motion regarding electromechanical interaction with simplifications ($g_\mathrm{eo} = g_\mathrm{om} = 0$ and $\Delta_\mu = \Delta_\mathrm{m} = 0$ where $\Delta_\mu$ and $\Delta_\mathrm{m}$ are detunings from microwave and mechanical resonance frequencies.),
\begin{equation}
    \begin{aligned}
        \dot{{c}} &= ig_{\mathrm{em}}{b} - \frac{\kappa_\mu}{2}{c}-\sqrt{\kappa_{\mu,\mathrm{,c}}} c_\mathrm{in}
        \\
        \dot{{b}} &= ig_{\mathrm{em}}{c} -\frac{\Gamma}{2}{b}
        .
    \end{aligned}
\end{equation}
Assuming steady state, number of piezoelectrically generated phonons can be expressed as
\begin{equation}
    \begin{aligned}
        n_\mathrm{m}
        &= \frac{4\kappa_{\mu\mathrm{,c}}}{\kappa_\mu^2}
        \left(\frac{g_\mathrm{em}}{\frac{2g_\mathrm{em}^2}{\kappa_\mu} + \frac{\Gamma}{2}}\right)^2 \langle c_\mathrm{in}^\dagger c_\mathrm{in}\rangle
        ,
    \end{aligned}
\end{equation}
where $\langle c_\mathrm{in}^\dagger c_\mathrm{in}\rangle$ denotes the input microwave drive flux ($\langle c_\mathrm{in}^\dagger c_\mathrm{in}\rangle=P_\mu/\hbar\Omega_\mu$) and $P_\mu$ is the input microwave drive power.
Considering the regime we are in ($\frac{2g_\mathrm{em}^2}{\kappa_\mu} \ll \frac{\Gamma}{2}$), above expression simplifies to
\begin{equation}\label{SIeq_nm}
    n_\mathrm{m}
    = \frac{4\kappa_{\mu\mathrm{,c}}}{\kappa_\mu^2}
    \left(\frac{2g_\mathrm{em}}{\Gamma}\right)^2 \langle c_\mathrm{in}^\dagger c_\mathrm{in}\rangle
    .
\end{equation}
In the experiment, $P_\mu$, measured with a spectrum analyzer, is swept from 4 dBm to 20dBm.
From Equation \ref{SIeq_nm}, we obtain electromechanical coupling rate, which turns out to be $g_\mathrm{em}/2\pi = 0.03$ Hz. Combining this with Equation \ref{eq:g_em}, we can correlate a bulk piezoelectric constant of 0.083 fm/V or a surface piezoelectric constant of 2.44 pm/V across a 1 nm thick piezoelectric layer.

\section{Alternative piezoelectricity measurement: Piezoresponse Force Microscopy (PFM)}

\begin{figure}[t]
    \centering
    \includegraphics[width=.7\textwidth]{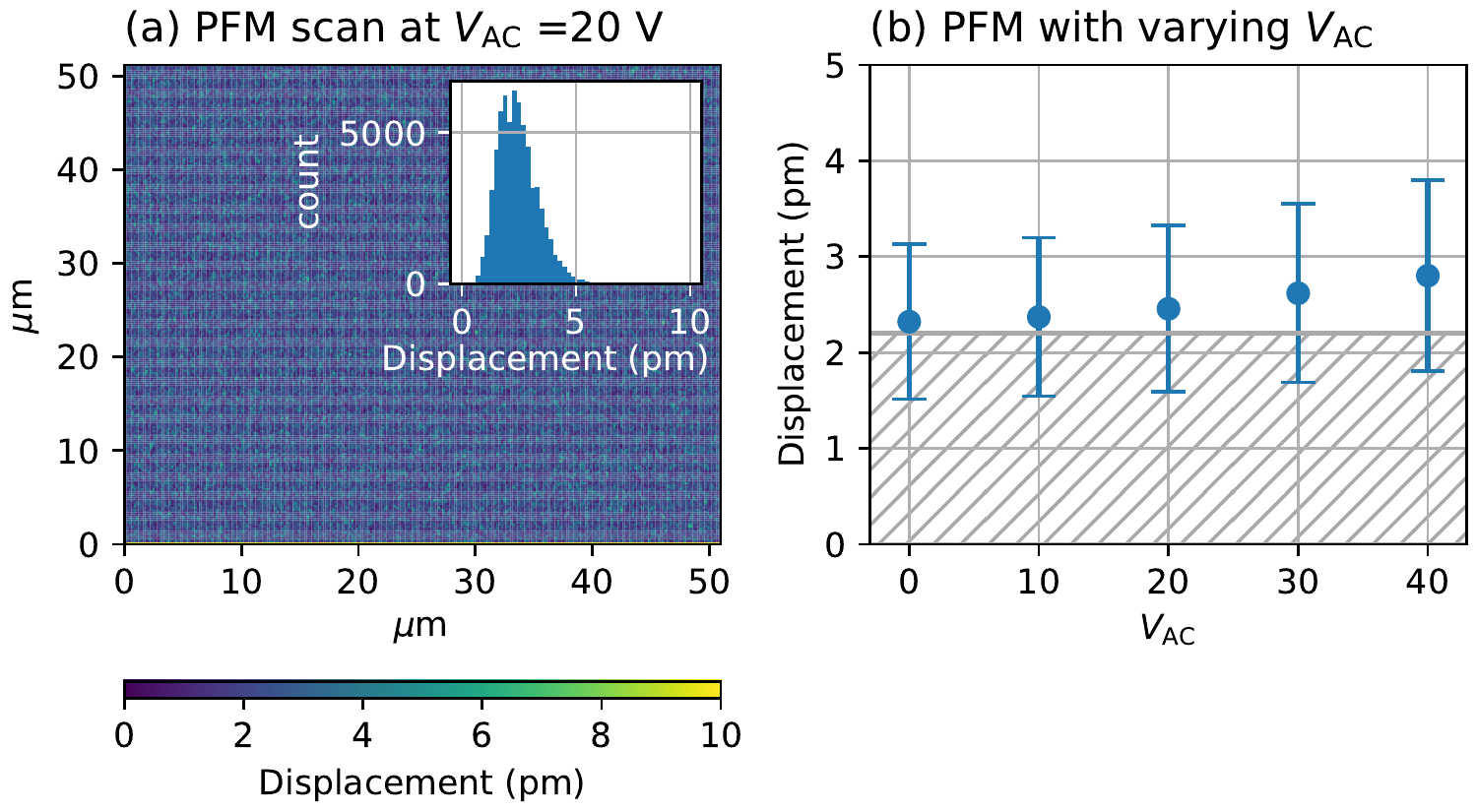}
    \caption{PFM scan across a 50 um$\times$50 um surface of CaF\textsubscript{2}. CaF\textsubscript{2} used here is the same substrate used in section 5 of the main text.(a) Example measurement at $V_\mathrm{AC}$ = 20 V. Histogram is shown in inset. (b) The average of displacement data is plotted with its standard deviation at various $V_\mathrm{AC}$.
    Displacement noise floor of $\sim$2.2 pm is indicated in dashed grey.
    Minimal variability in detected displacement indicates piezoelectricity smaller than the device's detection sensitivity.
    }
    \label{fig:SI_fig_PFM}
\end{figure}

Piezoresponse Force Microscopy (PFM) is a method of detecting piezoelectricity in $\sim$ 100 nm depth\cite{kelley_thickness_2020, collins_quantitative_2019, labuda_quantitative_2015}. The substrate is put under an AC electric field (usually in the range of $\sim$100 kHz), which leads to a piezoelectric deformation due to the converse piezoelectric effect. This deformation is mapped with a probe cantilever across the substrate. Recently an interferometric displacement sensing approach to PFM (IDS-PFM) has been developed which allows quantitative determination of out of plane electromechanical responses\cite{labuda_quantitative_2015, collins_quantitative_2019}. This method removes many of the artifacts that plagues traditional PFM, and is made inherently quantitative via the interferometric read out.

We put the same CaF\textsubscript{2} that we study in section 5 of the main text under a IDS-PFM as a way to alternatively detect piezoelectricity (see Figure \ref{fig:SI_fig_PFM}). The frequency of the AC field is $\sim$100 kHz. We measure a white noise floor of $\sim$70 fm/$\sqrt{\mathrm{Hz}}$, which is equivalent to a displacement noise floor of $\sim$2.2 pm in an imaging bandwidth of 1kHz. For a 1 V drive amplitude, this yields a minimum detectable piezo-sensitivity of $d_\mathrm{eff}$ = 2.2 pm/V. Minimum detectable piezo-sensitivity can be lowered by simply applying a larger drive voltage which was allowable in the case of CaF\textsubscript{2}, unlike thin films or materials having low dielectric breakdown potentials. For maximum applied voltage of 40 V, this implies a minimal detectable $d_\mathrm{eff}$ of 55 fm/V. From data, we barely see any deviation from the $\sim$2.2 pm background. Even though the data may appear to have a non-zero slope ($14 \pm 40$ fm/V), it is inconclusive that we detect any piezoelectricity since all data are within the standard deviation/error bar.
We attribute the smaller but finite slope of 14 fm/V to influence of electrostatic forces.
At high voltage especially, we can expect non-neglighble electrostatic forces and force gradients to act between the AFM tip and sample, which may lead to small but finite displacement of the tip position, even for the case of IDS-PFM (which is largely insensitive to cantilever-sample electrostatic effects\cite{labuda_quantitative_2015, collins_quantitative_2019}).
We further note that using softer (3 nN) cantilevers than the one used in Figure \ref{fig:SI_fig_PFM} ($\sim$42 nN) leads to a clear enlargement of the measured displacement (slope of $290 \pm 80$ fm/V) as would be expected for an electrostatic effect.

Therefore, we observe no detectable piezoelectricity though IDS-PFM, meaning that parasitic piezoelectricity in CaF\textsubscript{2} is smaller than the detection limit of 55 fm/V.
These results indicate the unlikely-hood of the parasitic piezoelectricity coming from an intrinsic surface layer, which requires piezoelectricity to be $\gtrsim$ 1pm/V.
Instead, it is more likely that the paracitic piezoelectricity detected in CaF\textsubscript{2} is originated from the bulk of the crystal.

\bibliographystyle{naturemag}
\bibliography{hybrid_v2.bib}